\documentclass[aps,prd,twocolumn,nofootinbib]{revtex4-1}
\usepackage{epsfig}
\usepackage[colorlinks,linkcolor=blue,anchorcolor=blue,citecolor=blue,urlcolor=blue,breaklinks=true]{hyperref}
\usepackage{graphics}
\usepackage{slashed}
\usepackage{color}
\usepackage{amsmath}
\usepackage{bm}
\usepackage{setspace}
\usepackage{caption}
\usepackage{multirow}

\begin{document}
\author{Cheng-Ming Li$^{1}$}
\author{Jin-Li Zhang$^{1}$}
\author{Yan Yan$^{2}$}
\author{Yong-Feng Huang$^{2}$}\email{hyf@nju.edu.cn}
\author{Hong-Shi Zong$^{1,3,4,}$}\email{zonghs@nju.edu.cn}
\address{$^{1}$ Department of Physics, Nanjing University, Nanjing 210093, China}
\address{$^{2}$ Department of Astronomy, Nanjing University, Nanjing 210093, China}
\address{$^{3}$Joint Center for Particle, Nuclear Physics and Cosmology, Nanjing 210093, China}
\address{$^{4}$State Key Laboratory of Theoretical Physics, Institute of Theoretical Physics, CAS, Beijing, 100190, China}
\title{Studies of the structure of massive hybrid stars within a modified NJL model}

\begin{abstract}
In this paper, we use the equation of state based on a modification of 2+1 flavors Nambu-Jona-Lasinio (NJL) model to study the quark matter of hybrid stars. For comparison, we utilize five EOSs of the relativistic mean-field (RMF) model to describe the hadronic phase. With the 3-window crossover interpolation approach, we try to construct relatively soft hybrid EOSs but find the maximum masses of hybrid stars do not differ too much. The results are quite close to two solar mass, which is consistent with the mass constraint of PSR J0348+0432. Furthermore it is noteworthy that the heaviest stable stars have central densities higher than that of the deconfinement transition thus suggesting a pure quark core in the hybrid star.

\bigskip

\noindent Key-words: equation of state, Nambu-Jona-Lasinio model, crossover, hybrid star
\bigskip

\noindent PACS Numbers: 12.38.Lg, 25.75.Nq, 21.65.Mn

\end{abstract}

\pacs{12.38.Mh, 12.39.-x, 25.75.Nq}

\maketitle

\section{INTRODUCTION}
The study of compact stars is a hot issue in astrophysics. Among many research directions, the study of the structure of hybrid stars is a popular topic. The large central density in the core and very low temperature make them natural laboratories to investigate strongly interacting matter within quantum chromodynamics (QCD). The interactions are so strong that perturbation theory can not be used here. On the other hand, the "sign problem" makes it difficult for lattice QCD (LQCD)~\cite{Borsanyi:2010bpa,PhysRevLett.110.172001} to deal with the calculations at finite baryon chemical potential in hybrid stars. Thus effective models are particularly useful in this regime, such as the Dyson-Schwinger equations (DSEs)~\cite{ROBERTS1994477,Roberts2000S1,doi:10.1142/S0218301303001326,Cloet20141,PhysRevD.90.114031,PhysRevD.91.034017,*PhysRevD.91.056003}, the quantum electrodynamics in 2+1 dimensions (QED$_3$)~\cite{ROBERTS1994477,PhysRevD.29.2423,PhysRevD.90.036007,PhysRevD.90.073013} and the Nambu-Jona-Lasinio (NJL) model~\cite{RevModPhys.64.649,Buballa2005205,Cui2013,NuclPhysB.896.682,doi:10.1142/S0217751X17500610}. In Refs.~\cite{PhysRevD.94.094001,BUBALLA200436,KLAHN2007170}, the authors considered the $2+1$ flavors NJL model to study quark matter in hybrid stars. However, the result of conventional NJL model does not match that of LQCD at zero chemical potential and finite temperature, which might cause some underlying problems when extending the calculation to finite chemical potential. To reproduce the lattice result, in Ref.~\cite{2016arXiv160807903F}, the authors introduced a modification of $2+1$ flavors NJL model with the four-fermion coupling strength dependent on the quark condensate, just like the operator product expansion (OPE) approach~\cite{Steele1989,PhysRevD.85.034031,Cui2013,PhysRevD.93.036006}. This modification makes the extension to finite chemical potential more reliable especially when studying the structure of hybrid stars.

For the region between the core and the crust of the hybrid star, a deconfinement phase transition (DPT) will happen. Despite the accurate depiction of dynamical chiral symmetry breaking (DCSB) by the NJL model (and its modified form), it lacks confinement and as a result the description of DPT becomes impossible. On one hand, the DPT and chiral phase transition (CPT) are two different kinds of phase transitions with different order parameters. On the other hand, in spite of good descriptions of both DPT and CPT by the Polyakov-loop extended NJL (PNJL) model at finite temperature and chemical potential, a series of studies~\cite{Ratti2007,PhysRevD.77.114028,Cui2014,PhysRevD.94.014008} consider the transitions to be crossover for DPT and first-order for CPT respectively. Furthermore, in the study of hybrid stars, the temperature is always fixed to zero resulting in a degeneration of the PNJL model with the conventional NJL model. The correlation between CPT and DPT is in fact still an open question with some authors thinking that there is a first-order DPT~\cite{PhysRevC.91.035803,Mallick201496,PhysRevD.91.034018} because of the mass twins~\cite{refId0} in the mass-radius (M-R) relation, while others~\cite{Alvarez-Castillo2015}, backed by the difficulty in the determination of the exact radius of a compact star, argue that for a smooth DPT mass twins can exist too. In addition, at low chemical potential and high temperature, the DPT is confirmed to be a crossover by LQCD~\cite{PhysRevLett.110.172001,PhysRevLett.113.152002,Endrodi:2015oba,Braguta:2015zta}, but at zero temperature and large chemical potential, there is still no model independent conclusion of the order of DPT. For the QCD phase diagram, whether there is a critical end point (CEP) and where it is are still unsolved theoretically. In general, the facts above leave the order of the DPT an open question, thus the study within a crossover DPT is meaningful in this regime. Recently, there are some papers utilizing the crossover DPT with different interpolation approaches to study massive hybrid stars and getting some good results~\cite{Masuda:2012ed,0004-637X-764-1-12,PhysRevD.91.045003,doi:10.1142/S0217732317500511,PhysRevD.95.056018}, that is, the hybrid stars with the maximum mass larger than two times of solar mass. Furthermore, in Ref.~\cite{PhysRevD.92.054012}, the authors demonstrate that there is not much difference on the maximum masses of the hybrid stars among different interpolation approaches.

We extend the study of the recent works~\cite{doi:10.1142/S0217732317501073,2016arXiv160807903F} and introduce a new EOS of quark matter. For comparison, we adopt five EOSs of hadronic matter in the relativistic mean-field (RMF) models to construct the hybrid EOSs. After solving the Tolman-Oppenheimer-Volkoff (TOV) equations, we get the M-R relation of the hybrid stars. The maximum masses obtained from the five hybrid EOSs are very similar, which are all about 2 times of the solar mass, thereby consistent with the mass constraint of PSR J0348+0432~\cite{Antoniadis1233232} and PSR J1614-2230~\cite{Fonseca:2016tux}. However, we can see that the hybrid EOSs are relatively soft via the diagram of sound velocities (namely the squared speed of sound). Additionally, the central densities of the heaviest stable stars are well beyond that of DPT, thus suggesting a pure quark core in the hybrid star.

The paper is organized as follows. In Sec.~\ref{one}, we present five RMF models~\cite{PhysRevC.55.540,PhysRevLett.86.5647,0067-0049-214-2-22,PhysRevC.71.024312,PhysRevC.76.045801,PhysRevC.81.034323,PhysRevC.94.035804} for the nuclear matter in hybrid stars and list some important properties in the table. In Sec.~\ref{two}, we introduce the new EOS of quark matter with the modification of $2+1$ flavors NJL model. In Sec.~\ref{three}, the 3-window interpolation approach~\cite{Masuda:2012ed,PhysRevC.93.035807,KOJO2016821} is adopted to construct the hybrid EOSs with a smooth DPT. The resulting M-R and M-$\rho_c$ ($\rho_c$ is the central density) relations are then obtained. For conclusion, we give a brief summary and discussion in Sec.~\ref{four}. Finally, the detailed derivation of the modified coupling constant $G$ is given in the Appendix~\ref{five}, which is inspired by QCD sum rule with OPE approach and satisfies the requirement of QCD in essence.

\section{EOS of hadronic matter}\label{one}
In this paper, we adopt some RMF nuclear models~\cite{PhysRevC.55.540,PhysRevLett.86.5647,0067-0049-214-2-22,PhysRevC.71.024312,PhysRevC.76.045801,PhysRevC.81.034323,PhysRevC.94.035804} to describe the confined hadronic matter system in the $\beta$-equilibrium, including: NL3~\cite{PhysRevC.55.540}, NL3$\omega\rho$~\cite{PhysRevLett.86.5647}, DD2~\cite{0067-0049-214-2-22}, DDME2~\cite{PhysRevC.71.024312} and BSR6~\cite{PhysRevC.76.045801,PhysRevC.81.034323}. The NL3, NL3$\omega\rho$ and BSR6 models belong to the nonlinear Walecka models (NLWM) which have constant coupling parameters. The other two models DD2 and DDME2 are the density-dependent models whose coupling parameters are density-dependent. There is a common property for the EOSs of these five models: they are all stiff enough to construct a neutron star heavier than two solar mass.

The unified EOS of hadronic matter in neutron stars is built in the following way: for the outer crust of the star we choose the Baym-Pethick-Sutherland EOS~\cite{1971ApJ...170..299B} to describe it, while for the inner crust and the core, the EOSs of the RMF models above are considered. Both the link from the outer crust to the inner crust and that from the inner crust to the core are smooth. None of the unified EOSs involve hyperons because, on one hand, the complete interactions between hyperons are still unknown and, on the other hand, the onset of hyperons for the unified EOSs occurs at densities larger than 0.28 fm$^{-3}$, already in the region of DPT and above the onset of quark matter.

Some critical properties of these unified hadronic EOS are demonstrated in Table~\ref{hadroniceos}, From the table, we can see that the maximum masses of the neutron stars are all heavier than 2.4 solar mass, well above 2 solar mass constraint implied by the astronomical observations of PSR J0348+0432~\cite{Antoniadis1233232}, PSR J1614-2230~\cite{Fonseca:2016tux}.
\begin{widetext}
\begin{center}
\begin{table}[h!]
\caption{Properties of some RMF models, including the Saturated density (n$_s$), energy per nucleon (E$_s$), compression modulus (K), symmetry energy (J), slope (L), the maximus mass for a purely nucleonic core composition (M$_{\odot}$) and the value of the onset densities of hyperons (n$_Y$).}\label{hadroniceos}
\begin{tabular}{p{1.9cm} p{1.5cm} p{1.5cm} p{1.5cm} p{1.5cm}p{1.5cm}p{1.5cm}p{1.1cm}}
\hline\hline
$Model$&$n_s$&$E_s$&$K$&$J$&$L$&$M$&$n_{Y}$\\
\,&(fm$^{-3}$)&(MeV)&(MeV)&(MeV)&(MeV)&(M{$_{\odot}$})&(fm$^{-3}$)\\
\hline
NL3&0.149&-16.2&271.6&37.4&118.9&2.77&0.28\\
\hline
NL3$\omega\rho$&0.148&-16.2&271.6&31.7&55.5&2.75&0.31\\
\hline
DD2&0.149&-16.0&242.6&31.7&55.0&2.42&0.37\\
\hline
DDME2&0.152&-16.1&250.9&32.3&51.2&2.48&0.34\\
\hline
BSR6&0.149&-16.1&235.8&35.6&85.7&2.44&0.33\\
\hline\hline
\end{tabular}
\end{table}
\end{center}
\end{widetext}

\section{EOS of quark matter}\label{two}
The general form of the Lagrangian of $2+1$ flavors NJL model is
\begin{eqnarray}
\mathcal{L}=&&\bar{\psi}(i{\not\!\partial}-m)\psi+G[(\bar{\psi}\lambda_{i}\psi)^2+(\bar{\psi}i\gamma_{5}\lambda_{i}\psi)^2]\nonumber\\
&&-K\,({\rm det}[\bar{\psi}(1+\gamma_{5})\psi]+{\rm det}[\bar{\psi}(1-\gamma_{5})\psi]),\,\,\label{lagrangian}
\end{eqnarray}
where $G$ and $K$ are the four-fermion and six-fermion interaction coupling constant, respectively; $\lambda^{\rm a}, {\rm a}=1\rightarrow8$ is the Gell-Mann matrix, and $\lambda^0$ is defined as $\sqrt{\frac{2}{3}}\,I$ ($I$ is the identity matrix). Then the mean-field thermodynamic potential reads
\begin{eqnarray}
  \Omega(T,\{\mu_f\},\{\phi_f\})=&&\sum_{f=u,d,s}\Omega_{M_f}(T,\mu_f)+2G(\phi^2_u+\phi^2_d+\phi^2_s)\nonumber\\
  &&-4K\phi_u \phi_d \phi_s+{\rm const}.\label{thermopot}
\end{eqnarray}
Here $\Omega_{M_f}$ is the contribution of the gas of quasiparticles of flavor f, and $\phi_f$ is the quark condensate of flavor f, $f=u,d,s$.
\begin{eqnarray}
  \Omega_{M_f}=&&-2N_c\int\frac{d^3 p}{(2\pi)^3}\{T\,{\rm ln}(1+{\rm exp}(-\frac{1}{T}(E_{p.f}-\mu_f)))\nonumber\\
  &&+T\,{\rm ln}(1+{\rm exp}(-\frac{1}{T}(E_{p,f}+\mu_f)))+E_{p,f}\}.\label{thermopotofquasi}
\end{eqnarray}
In the Hartree approximation, the corresponding gap equation is now given by
\begin{eqnarray}
  M_i &=& m_i-4G\phi_i+2K\phi_j \phi_k,\nonumber\\
  && (i,j,k)=any\,permutation\,of\,(u,d,s).\label{gap}
\end{eqnarray}
In the treatment of thermodynamically consistency, for flavor i, the quark condensate $\phi_i$ and the particle number density $\rho_i$ should follow from $\Omega$ as
\begin{eqnarray}
  \phi_i &=& \frac{\partial \Omega}{\partial m_i}\nonumber \\
         &=& -2 N_c\int\frac{d^3 p}{(2\pi)^3}\frac{M_i}{E_{p,i}}[1-n_{p,i}(T,\mu_i)-\overline{n}_{p,i}(T,\mu_i)],\nonumber\\
         \label{phi} \\
  \rho &=& -\frac{\partial \Omega}{\partial \mu_i}\nonumber \\
       &=& 2 N_c\int\frac{d^3 p}{(2\pi)^3}(n_{p,i}(T,\mu_i)-\overline{n}_{p,i}(T,\mu_i)),\label{rho}
\end{eqnarray}
here $E_p=\sqrt{\overrightarrow{p}^2+M^2}$ is the on-shell energy of the quark, while $n_{p,i}(T,\mu_i)$, $\overline{n}_{p,i}(T,\mu_i)$ are the Fermi occupation numbers of quarks and antiquarks of flavor i respectively, which are defined as
\begin{eqnarray}
  n_{p,i}(T,\mu_i) &=& [{\rm exp}^{(E_p-\mu_i)/T}+1]^{-1},\label{fonofq} \\
  \overline{n}_{p,i}(T,\mu_i) &=& [{\rm exp}^{(E_p+\mu_i)/T}+1]^{-1}.\label{fonofantiq}
\end{eqnarray}

It is known that NJL model can not be renormalized because of the four-fermion and six-fermion interaction in the Lagrangian. Thus, for terms containing divergent integrals, a certain regularization method has to be applied. In a non-renormalizable model, such as NJL model, one has a choice: a cutoff (or cutoffs) can be placed on all integrals~\cite{PhysRevD.75.065004,PhysRevD.81.016007,doi:10.1142/S0217751X12500601,BRATOVIC2013131} or, a
cutoff can be placed on only those integrals which are ultraviolet divergent~\cite{FUKUSHIMA2004277,PhysRevD.84.014011,PhysRevD.83.105008,PhysRevD.85.054013,PhysRevD.75.034007}. However, just as pointed out by Ref.~\cite{PhysRevD.94.071503} that the best choice of regularization scheme is whether it
treats all momentum integrals equally, viz. if a cutoff is placed on one, then it should be placed on all. Following this perspective, we use the three-momentum cutoff to regularize all the integrals in the manuscript. Before further calculation, we need to fix five parameters to fit five experimental data at zero temperature and chemical potential. This procedure is similar to that used in Ref.~\cite{HATSUDA1994221}.

In this paper, we will use a modified $2+1$ flavors NJL model,
with the four-point coupling constant being dependent on the
quark condensate, inspired by the OPE method. Due to the ``sign problem``
of lattice QCD simulation at finite chemical potential, lattice QCD
method is limited to low baryon chemical potential and
finite temperature right now and conventional NJL model
can not match lattice results at zero chemical potential
and finite temperature. However, just as we will show
below, with a coupling strength depending on the quark
condensate, we are able to reproduce the lattice result at finite temperature and zero chemical potential. Here, a natural question arises: is our modification of the usual NJL model reasonable? We try to answer this question from two aspects.
Firstly, It is well known that the quark propagator and gluon propagator satisfy their respective DSEs and they are coupled with each other by QCD in essence. The quark propagator in Nambu and Wigner phase are very different. Hence, the corresponding gluon propagators in these two phases should be different, too. However, most work within NJL model don't meet this requirement (In the framework of NJL model, the coupling constant $G$ has a physical meaning of the effective gluon propagator. In normal NJL model, $G$ is the same constant whether it is in Nambu or Wigner phase). Our manuscript incorporates quark's feedback into gluon propagator and therefore provides a method to fulfill this requirement by QCD. Secondly, in the normal NJL model, the effective gluon propagator, namely, the coupling constant $G$ does not change with temperature and chemical potential. However, Lattice QCD simulation has shown that the gluon propagator evolves with temperature, although its evolution with chemical potential is unknown. This means that the "static" effective gluon propagator $G$ used by most NJL studies is unreasonable in essence. Nevertheless, in our manuscript the coupling $G$ being dependent on the quark condensate will change naturally with temperature and chemical potential, since the quark condensate is temperature and chemical potential dependent.
In fact, to resolve this problem, some papers~\cite{FISCHER20131036,FISCHER2011438} propose to consider the contribution of quark loops in the DSE of the gluon to incorporate the feedback of quark into the gluon propagator. However, the analysis and fitting of the lattice data is critical for the quenched part of gluon propagator. In this work, we investigate an alternative treatment in the OPE framework, and the specific implementation of how to extract the feedback of quark from gluon propagator can be found in the Appendix~\ref{five}. From this derivation, the coupling constant $G$ in the conventional $2+1$ flavors NJL model can be modified as
\begin{equation}\label{couplingstrength}
  G\rightarrow G_1+G_2(\phi_u+\phi_d+\phi_s).
\end{equation}

Then, to determine the ratio of $G_1$ to $G_2$, we fit the critical temperature $T_c$ to the result of lattice QCD at zero chemical potential, which is about 158 MeV. What's more, our calculation for the light quark condensate $\langle\bar{\psi}\psi\rangle_R$ and the subtracted chiral condensate $\Delta_{l,s}$ at zero chemical potential coincides with the results of lattice QCD~\cite{Borsanyi2010}. The quantities above are defined in Ref.~\cite{Borsanyi2010}: $\langle\bar{\psi}\psi\rangle_R=-[\langle\bar{\psi}\psi\rangle_{l,T}-\langle\bar{\psi}\psi\rangle_{l,0}]\frac{m_l}{X^4}$, $\Delta_{l,s}=\frac{\langle\bar{\psi}\psi\rangle_{l,T}-\frac{m_l}{m_s}\langle\bar{\psi}\psi\rangle_{s,T}}{\langle\bar{\psi}\psi\rangle_{l,0}-\frac{m_l}{m_s}\langle\bar{\psi}\psi\rangle_{s,0}}$.
Here the subscript $l$ represents $u$ or $d$ quark, while $s$ represents $s$ quark. $X$ can be any quantity with a dimension of mass, here we choose it as $m_{\pi}$. The results are presented in Fig.~\ref{Fig:latticecompare} showing that, with only one more parameter, we are able to fit the lattice results at finite temperature quite well. From this figure, we can also find that at zero temperature and chemical potential, our modified NJL model will return to the normal NJL model, thus leading to many important results and properties of the vacuum unaffected by our modification. The whole parameter set is shown in Table~\ref{parameters}.
\begin{widetext}
\begin{center}
\begin{figure}
\includegraphics[width=0.87\textwidth]{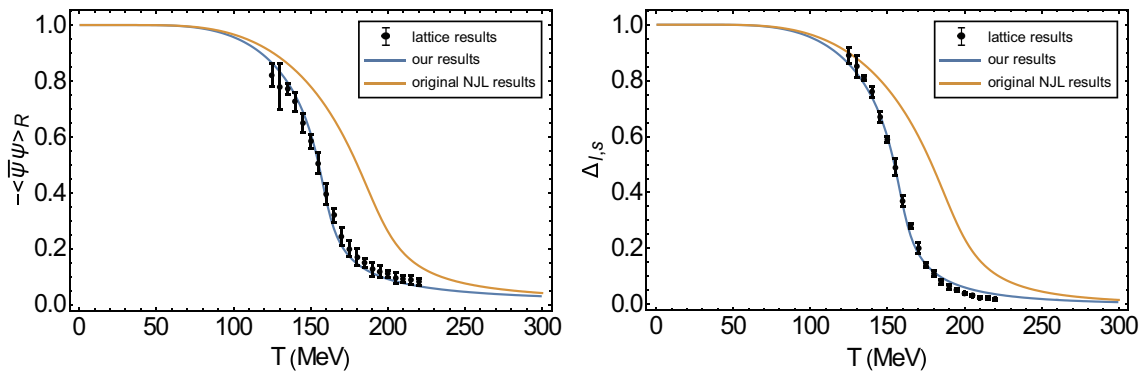}
\caption{Left: the temperature dependence of light quark ($u$, $d$) condensate (normalized to their vacuum value), compared with the lattice result in Ref.~\cite{Borsanyi:2010bpa}. Right: the temperature dependence of $\Delta_{l,s}$ (a linear combination of $\phi_u$ and $\phi_s$).}
\label{Fig:latticecompare}
\end{figure}
\end{center}
\end{widetext}

\begin{widetext}
\begin{center}
\begin{table}[h!]
\caption{Parameter set fixed for the quark matter.}\label{parameters}
\begin{tabular}{p{1.6cm} p{1.6cm} p{2cm} p{2cm} p{2cm} p{1.6cm}}
\hline\hline
$m_u$\,(MeV)&$m_s$\,(MeV)&$G_1$\,(MeV$^{-2}$)&$G_2$\,(MeV$^{-5}$)&$K$\,(MeV$^{-5}$)&$\Lambda_{UV}$(MeV)\\
\hline
5&136&3.74$\times10^{-6}$&-1.74$\times10^{-14}$&9.29$\times10^{-14}$&631\\
\hline\hline
\end{tabular}
\end{table}
\end{center}
\end{widetext}
\begin{figure}
\includegraphics[width=0.47\textwidth]{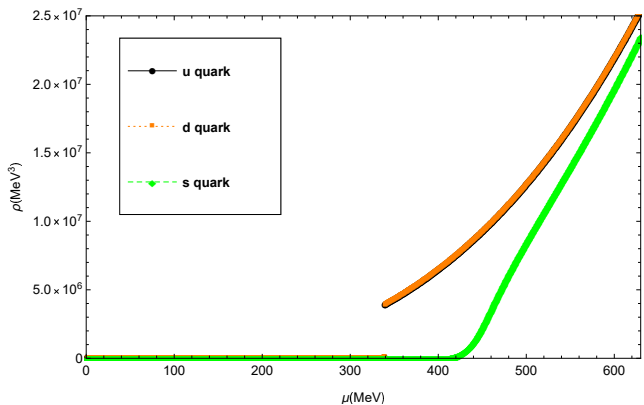}
\caption{The relations between $u$, $d$ and $s$ quark number densities and the chemical potential which are shown as the black line, orange line and green line, respectively. For $\mu<$340 MeV, the three lines coincide and are close to X-axis, while for $\mu>$340 MeV, only black line and orange line coincide.}
\label{Fig:nurelation}
\end{figure}

Here, it is interesting to look into the connection between our modified model and those with multi-quark interactions, for example, the NJL model with six-quark and eight-quark interactions~\cite{PhysRevD.81.116005,OSIPOV200648,OSIPOV20072021}. If we focus on the mass gap of the quark in the conventional $2+1$ flavors NJL model, we can find that its linear term $-4G\phi_i$ plays an important role in DCSB. Both of our modified model and those with multi-quark interactions give a modification on the term. In our manuscript, the result of the revision of the term $-4G\phi_i$ is $-4\left[G_1+G_2(\phi_u+\phi_d+\phi_s)\right]\phi_i$, while in the modified NJL model with multi-quark interactions ~\cite{PhysRevD.81.116005,OSIPOV200648,OSIPOV20072021}, the corresponding correction result is $-\left[G+g_1(h_u^2+h_d^2+h_s^2)/4\right]h_i$ (where $h_i$ is defined as $\phi_i=\left((h_i)\big|_{\Delta_i\neq 0}-(h_i)\big|_{\Delta_i=0}\right)/2$, $\Delta_i$ is the mass gap of flavor i).
Nevertheless, because of the very low temperature in hybrid stars, the critical baryon chemical potential is always supposed to be about 1-1.2 GeV~\cite{PhysRevD.77.114028,0034-4885-74-1-014001}. Thus in this case, as the condensates are expected to be small in this regime, the influence of multi-quark interactions, especially the quadratic modification stemming from eight quark interactions, can be expected to be weak, leading to a small difference of the linear term between the conventional $2+1$ flavors NJL model and the modification with multi-quark interactions. This is consistent with the result of our discussion in the
Appendix VI if we incorporate the contribution of multi-quark interactions into
that of $-4G_1\phi_i$..


Now we can extend our calculation to finite temperature and chemical potential. Because the temperature of hybrid stars is quite low, we take it as $10^{-5}$ MeV for numeral calculations in this paper. The relations between the particle number densities and the chemical potentials of $u$, $d$ and $s$ quark are shown in Fig.~\ref{Fig:nurelation}. We can see that the chemical potential dependence of the density of $u$ quark is same to that of $d$ quark because of the isospin symmetry between $u$ and $d$ quark. And their critical chemical potentials are about 340 MeV, where a gap appears. After this point, the densities increase monotonically. But for $s$ quark, the critical chemical potential is about 410 MeV where its density starts to be nonzero. Besides, its density curve is smooth all the time. Thus we can conclude that $u$ ($d$) quark undergoes a first-order CPT, while $s$ quark undergoes a crossover CPT as the chemical potential rises. In the following, we will consider the electro-weak interaction, and the constraints of chemical equilibrium as well as electric charge neutrality ($\beta$-equilibrium),
\begin{equation}\label{constrains}
  \left\{\begin{array}{lcl}
           \mu_{\rm d}=\mu_{\rm u}+\mu_{\rm e}, \\
           \mu_{\rm s}=\mu_{\rm u}+\mu_{\rm e}, \\
           \frac{2}{3}\rho_{\rm u}=\frac{1}{3}\rho_{\rm d}+\frac{1}{3}\rho_{\rm s}+\rho_{\rm e}.
         \end{array}\right.
\end{equation}
\begin{figure}
\includegraphics[width=0.47\textwidth]{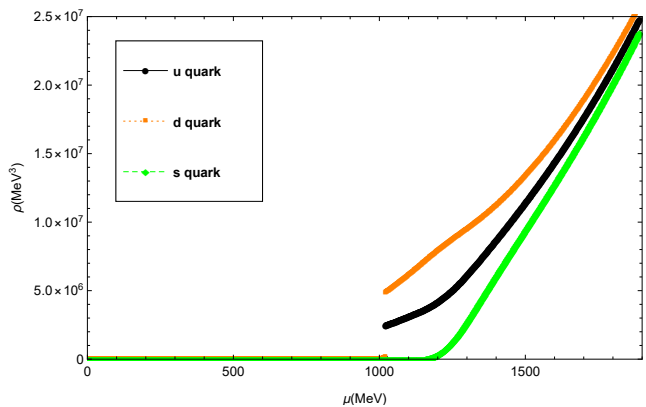}
\caption{Considering the electric charge neutrality conditions and the chemical equilibrium, the relations between $u$, $d$ and $s$ quark number densities and the chemical potential are depicted as the black line, orange line and green line, respectively. For $\mu<$340 MeV, the three lines coincide and are close to X-axis.}
\label{Fig:nurelationequ}
\end{figure}
Then we can deduce the relations between the densities of the constituent quarks ($u$, $d$ and $s$) and the baryon chemical potential of the equilibrium system, which is shown in Fig.~\ref{Fig:nurelationequ}.

By definition, the pressure of the system at zero temperature is~\cite{doi:10.1142/S0217751X08040457}
\begin{equation}\label{EOSofQCD}
  P(\mu)=P(\mu=0)+\int_{0}^{\mu}d\mu'\rho(\mu').
\end{equation}
Here $P(\mu=0)$ is pressure of the vacuum, standing for the negative pressure of the vacuum at zero chemical potential, which is treated as a model-dependent phenomenological parameter. The confinement of QCD is therefore included in the same sense as it is present in the MIT bag model. Before the following calculation, we have to ascertain its value. Like Ref.~\cite{PhysRevD.92.054012}, we associate $P(\mu=0)$ with $-B$ (the vacuum bag constant). It's well known that $B$ is a phenomenological parameter in the range of (100 MeV)$^4$-(200 MeV)$^4$~\cite{PhysRevD.46.3211,LU1998443}. Recently, based on the observation of pulsar and the recent binary neutron star (BNS) merger event GW170817, Ref.~\cite{Zhou:2017pha} gives a quite narrow limitation to $B$,
that is, from (134.1 MeV)$^4$ to (141.4 MeV)$^4$. So in this work we choose the middle value of the range and fix $B=(137$ MeV$)^4$. (The study within the regime of $B=(100$ MeV$)^4$ has also been implemented, and the relevant results are shown in the end of Sec.~\ref{three}.)

Now we can get the relation between the energy density of the quark matter and its pressure~\cite{PhysRevD.86.114028,PhysRevD.51.1989}
\begin{equation}\label{rbedasp}
  \epsilon=-P+\sum_{i}\mu_{\rm i}\rho_{\rm i}.
\end{equation}

\section{Structure of hybrid stars}\label{three}
To obtain the hybrid EOS with a smooth DPT. We need to employ an appropriate interpolation approach to connect hadronic matter phase and quark matter phase. In Ref.~\cite{Masuda:2012ed,PhysRevC.93.035807,KOJO2016821}, the authors employ a 3-window modeling of QCD matter. Specifically, it is the P-interpolation and $\epsilon$-interpolation approaches in the P-$\rho$ and $\epsilon$-$\rho$ plane, respectively. As an extension, the authors of Ref.~\cite{PhysRevD.92.054012} use the P-interpolation but in the P-$\mu$ plane. In this paper, we will adopt the same interpolation method as the Refs.~\cite{PhysRevD.92.054012,PhysRevD.95.056018}. And the interpolating process is defined in the following,
\begin{eqnarray}
  P(\mu) &=& P_{\rm H}(\mu)f_-(\mu)+P_{\rm Q}(\mu)f_+(\mu)\,\,,\nonumber\\
  f_{\pm}(\mu) &=& \frac{1}{2}(1\pm {\rm tanh}\,(\frac{\mu-\bar{\mu}}{\Gamma}))\,\,.\label{interpolation}
\end{eqnarray}
\begin{figure}
\includegraphics[width=0.47\textwidth]{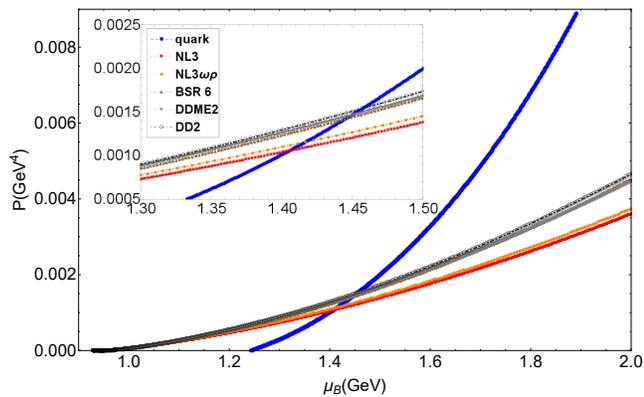}
\caption{The baryon chemical potential dependence of pressure for different matters (blue dashed line, red dotted line, orange dotted line, brown dot-dashed line, gray dot-dashed line and black dot-dashed line correspond to quark matter and hadronic matter of NL3, NL3$\omega\rho$, BSR6, DDME2, DD2 models respectively).}
\label{Fig:quarkandhadronEOS}
\end{figure}
\begin{figure}
\includegraphics[width=0.47\textwidth]{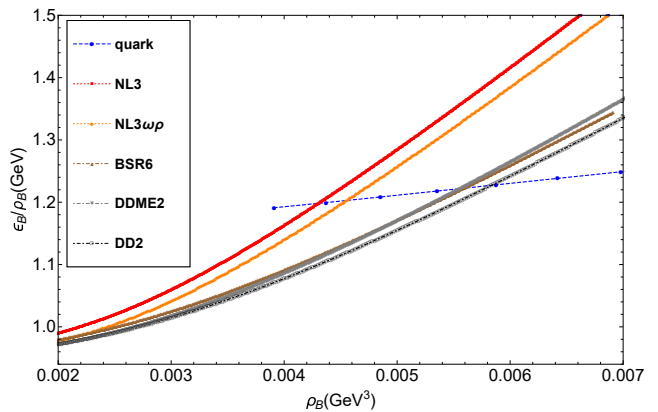}
\caption{The energy per baryon vs the baryon particle number density for different matters (blue dashed line, red dotted line, orange dotted line, brown dot-dashed line, gray dot-dashed line and black dot-dashed line correspond to quark matter and hadronic matter of NL3, NL3$\omega\rho$, BSR6, DDME2, DD2 models respectively).}
\label{Fig:bindingenergy}
\end{figure}
\begin{figure}
\includegraphics[width=0.47\textwidth]{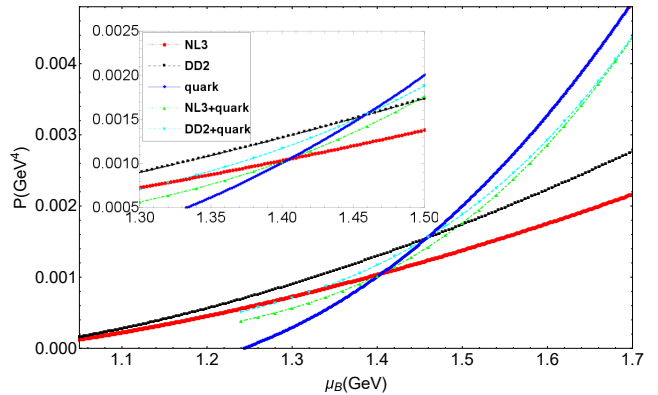}
\caption{EOSs of hadronic matter of NL3, DD2 models (red dot-dashed line, black dashed line respectively), EOS of quark matter (blue real line), hybrid EOSs constructed by NL3, DD2 models and the modified NJL model (green dot-dashed line, cyan dashed line respectively).}
\label{Fig:hybridEOS}
\end{figure}
\begin{figure}
\includegraphics[width=0.47\textwidth]{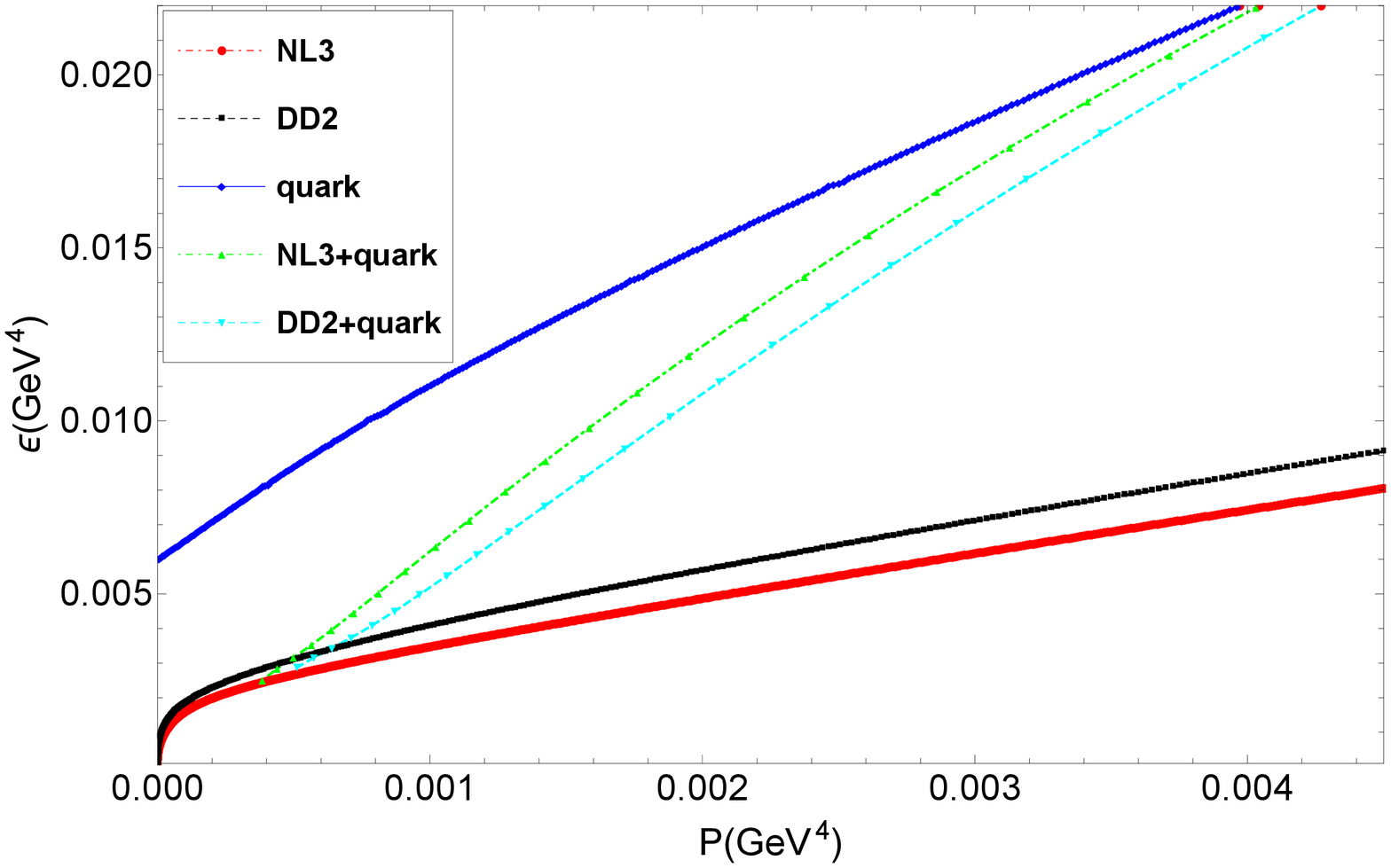}
\caption{$\epsilon$-P relations of hadronic matter of NL3, DD2 models (red dot-dashed line, black dashed line respectively), $\epsilon$-P relation of quark matter (blue real line), $\epsilon$-P relations of the hybrid system constructed by NL3, DD2 models and the modified NJL model (green dot-dashed line, cyan dashed line respectively).}
\label{Fig:eprelation}
\end{figure}
\begin{figure}
\includegraphics[width=0.47\textwidth]{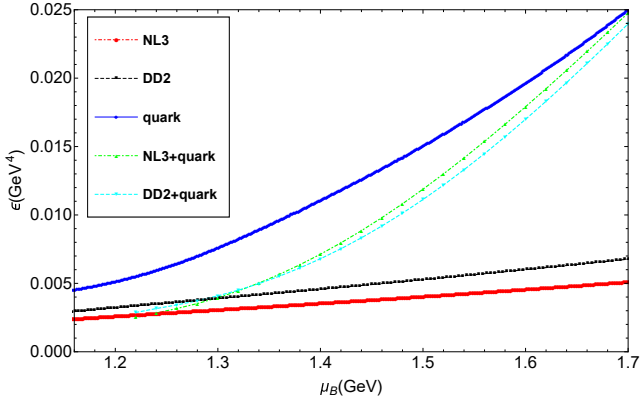}
\caption{$\epsilon$-$\mu$ relations of hadronic matter of NL3, DD2 models (red dot-dashed line, black dashed line respectively), $\epsilon$-$\mu$ relation of quark matter (blue real line), $\epsilon$-$\mu$ relations of the hybrid system constructed by NL3, DD2 models and the modified NJL model (green dot-dashed line, cyan dashed line respectively).}
\label{Fig:emurelation}
\end{figure}
Here $P_H$ and $P_Q$ denote the pressures of hadronic matter and quark matter respectively. And the association factors $f_{\pm}$ are the interpolation functions to achieve a crossover in the region of DPT. The range of the region is characterized by the window of the function, $\bar{\mu}-\Gamma\lesssim\mu\lesssim\bar{\mu}+\Gamma$, where hadrons coexist and strongly interact with quarks. Thus neither the EOS of hadron matter nor the EOS of quark matter can play a role alone. The EOS of quark matter as well as the EOSs of hadronic matter are shown in Fig.~\ref{Fig:quarkandhadronEOS}. We can find that the chemical potentials of the intersections of quark EOS and hadronic EOSs are from 1.40 GeV to 1.45 GeV, corresponding to the center points of the DPT region. Considering the critical baryon chemical potential of DPT at zero temperature, which is predicted to be larger than 1 GeV~\cite{PhysRevD.77.114028,0034-4885-74-1-014001}, the parameter $\Gamma$ should be no larger than 0.4 GeV. As it is shown in Ref.~\cite{Masuda:2012ed}, a larger window corresponds to a softer hybrid EOS. From the viewpoint of constructing a relatively soft hybrid EOS, we choose $\Gamma=0.4$ GeV. Thus the whole region of DPT should be located at the chemical potential interval from about 1 GeV to about 1.8 GeV. According to the saturated nuclear densities of different hadronic EOSs demonstrated in Table.~\ref{hadroniceos}, the lower boundaries of the region correspond to about $\rho_0^j$ ($j$ denotes the species of the hadronic EOS) while the upper boundaries are 4.03 $\rho_0^{NL3}$, 4.23 $\rho_0^{NL3\omega\rho}$, 5.53 $\rho_0^{DD2}$, 5.13 $\rho_0^{DDME2}$, 5.41 $\rho_0^{BSR6}$ respectively. By comparison, the five EOSs of hadronic matter can be divided into two categories for their similarity, that is, NL3 and NL3$\omega\rho$ for one category, DD2, DDME2 and BSR6 for the other category. To compare the stabilities of the quark matter and hadronic matter, we calculate the binding energy of them and the result is shown in Fig.~\ref{Fig:bindingenergy}. There are five intersections of quark matter and hadronic matter in this figure, corresponding to the particle number densities of 3.75 $\rho_0^{NL3}$, 4.0 $\rho_0^{NL3\omega\rho}$, 5.09 $\rho_0^{DD2}$, 4.75 $\rho_0^{DDME2}$, 4.86 $\rho_0^{BSR6}$ respectively. When $\rho_B$ is smaller than these intersections, the hadronic matter is more stable than quark matter. But when $\rho_B$ is larger than these points of intersection, the quark matter is more stable. What's more, these intersections are all in the range of their own interpolating windows, which confirms the validity of our interpolation approach.

Now we can obtain the five hybrid EOSs. But for a more sharp contrast, we will enumerate the hadronic EOSs of NL3 and DD2 to represent the EOSs of the two categories in the following. In Fig.~\ref{Fig:hybridEOS}, we can see the hybrid EOSs of the two categories. In the region of high chemical potential, the hybrid EOSs are very similar because of their asymptotic behaviors to the EOS of quark matter. However, in the region of low chemical potential, they are a little different on account of their approaching to different EOSs of hadronic matter. Then we deduce the pressure dependence of the energy density and the chemical potential dependence of the energy density for pure quark matter, pure hadronic matter and the hybrid matter respectively, and the results are shown in Fig.~\ref{Fig:eprelation} and Fig.~\ref{Fig:emurelation}. The asymptotic behaviors of the curves of hybrid matter in these two pictures are analogous to that in Fig.~\ref{Fig:hybridEOS}. As we can see, the property of the interpolation method make the hybrid EOSs smooth curves, which are close to the hadronic EOS at small baryon chemical potential, but close to the EOS of quark matter at large baryon chemical potential.

We also calculate the sound velocities (the squared speed of sound) of the five hybrid EOSs to investigate the stiffness of them. By definition, the sound velocity of a certain system is
\begin{equation}\label{soundvelocity}
 \nu_{\rm s} = \sqrt{\frac{{\rm d}p}{{\rm d}\epsilon}}.
\end{equation}
\begin{figure}
\includegraphics[width=0.47\textwidth]{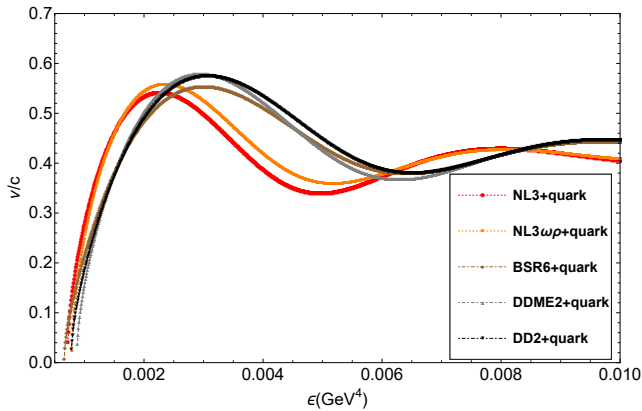}
\caption{The sound velocities of the five hybrid systems constructed by the NL3, NL3$\omega\rho$, BSR6, DDME2, DD2 hadronic models and the modified NJL model which are depicted in red dotted line, orange dotted line, brown dot-dashed line, gray dot-dashed line, black dot-dashed line, respectively.}
\label{Fig:soundvelocity}
\end{figure}
\begin{figure}
\includegraphics[width=0.47\textwidth]{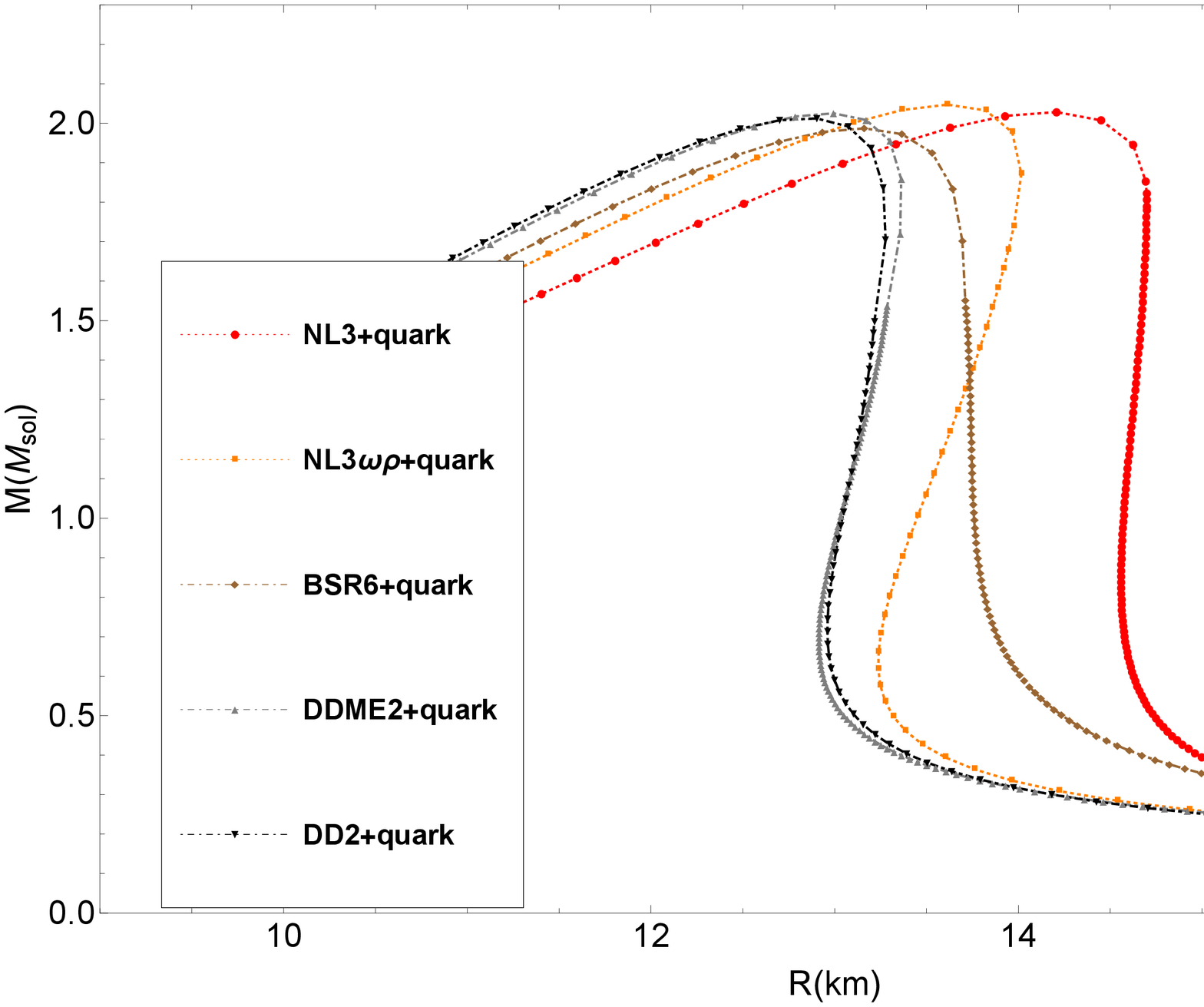}
\caption{The M-R relations of the five hybrid systems constructed by the NL3, NL3$\omega\rho$, BSR6, DDME2, DD2 hadronic models and the modified NJL model which are depicted in red dotted line, orange dotted line, brown dot-dashed line, gray dot-dashed line, black dot-dashed line, respectively.}
\label{Fig:mrrelation}
\end{figure}
\begin{figure}
\includegraphics[width=0.47\textwidth]{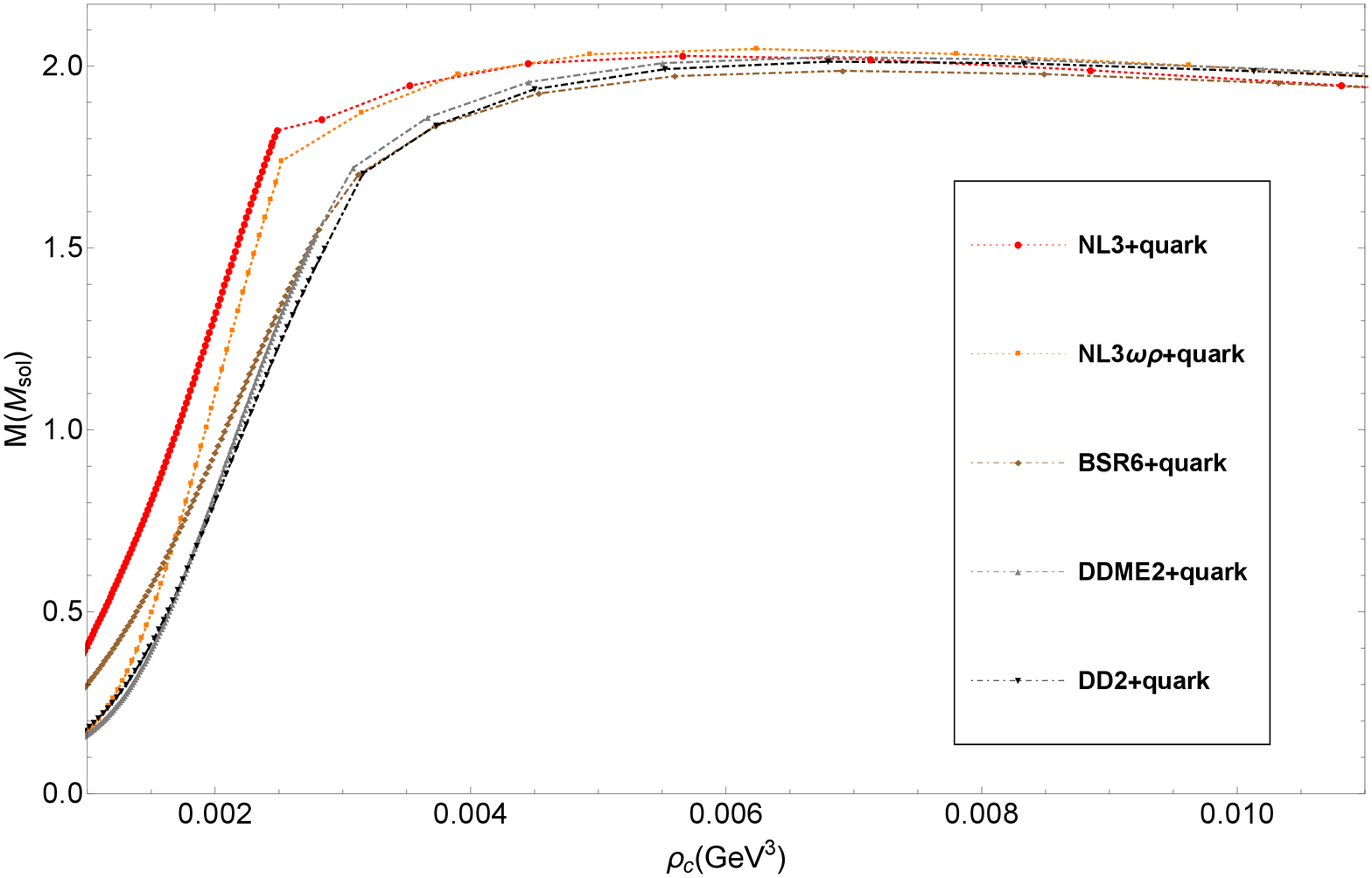}
\caption{The M-$\rho_c$ relations of the five hybrid systems constructed by the NL3, NL3$\omega\rho$, BSR6, DDME2, DD2 hadronic models and the modified NJL model which are depicted in red dotted line, orange dotted line, brown dot-dashed line, gray dot-dashed line, black dot-dashed line, respectively.}
\label{Fig:mrhorelation}
\end{figure}
In Fig.~\ref{Fig:soundvelocity}, the sound velocities of the hybrid EOSs are all smaller than 0.6 times of light speed despite the different trends of them, which demonstrates that our hybrid EOSs are reasonable and relatively soft.

To get the configuration of the hybrid stars, we integrate the TOV equations with the five hybrid EOSs above. The results of M-R relation and M-$\rho_c$ relation are shown in Fig.~\ref{Fig:mrrelation} and Fig.~\ref{Fig:mrhorelation} respectively, here $\rho_c$ represents the central density of the hybrid star. We can see that the maximum mass of the hybrid stars is 2.03, 2.05, 2.01, 2.03, 1.99 times of solar mass with the radii of 14.21, 13.61, 12.90, 12.99, 13.16 kilometers, respectively. Considering the recent astronomical observations of massive neutron stars, PSR J0348+0432 and PSR J1614-2230 with the mass constraint of 2.01$\pm$0.04 M$_\odot$ and 1.97$\pm$0.04 M$_\odot$, are two of the best massive neutron star observations until now. Obviously, our result is in good accordance with the mass constraint. And the central densities of the heaviest stable stars are also larger than that of their corresponding DPT, reaching to 5.15 $\rho_0^{NL3}$, 5.30 $\rho_0^{NL3\omega\rho}$, 6.33 $\rho_0^{DD2}$, 6.04 $\rho_0^{DDME2}$, 5.98 $\rho_0^{BSR6}$ respectively, which confirms a pure quark core in the hybrid star. For comparison, we also calculate the hybrid EOSs, the M-R relation and the M-$\rho_c$ relation in the regime of $B=($100 MeV$)^4$. The results are shown in Table~\ref{hybridresults}, indicating that there is not widely difference between the result of two regimes, ($B=($100 MeV$)^4$ and $B=($137 MeV$)^4$). When $B=($100 MeV$)^4$, the hybrid EOSs are soft too, and the maximum mass of the hybrid star is 1.98, 2.02, 2.00, 2.01, 1.97 times of solar mass with the radii of 13.93, 13.51, 12.80, 12.97, 13.13 kilometers, respectively.
\begin{table}[h!]
\caption{The results of the maximum mass $M_{max}$, the radii $R_m$, the central density $\rho_c$ of the heaviest stable hybrid stars for the two bag constants: $B=($100 MeV)$^4$ and $B=($137 MeV)$^4$. (Here M$_{\bigodot}$ represents one solar mass.)}\label{hybridresults}
\begin{tabular}{p{3cm} p{1.3cm} p{1.3cm} p{1.3cm} p{1.3cm}}
\hline\hline
\multirow{2}{*}{EOSs}&$B$&$M_{max}$&$R_m$&$\rho_c$\\
&(MeV$^4$)&(M$_{\bigodot}$)&(km)&(GeV$^3$)\\
\hline
\multirow{2}{*}{NL3+quark}&100$^4$&1.98&13.93&0.00625\\
&137$^4$&2.03&14.21&0.00590\\
\hline
\multirow{2}{*}{NL3$\omega\rho$+quark}&100$^4$&2.02&13.51&0.00648\\
&137$^4$&2.05&13.61&0.00602\\
\hline
\multirow{2}{*}{DD2+quark}&100$^4$&2.00&12.80&0.00713\\
&137$^4$&2.01&12.90&0.00723\\
\hline
\multirow{2}{*}{DDME2+quark}&100$^4$&2.01&12.97&0.00711\\
&137$^4$&2.03&12.99&0.00706\\
\hline
\multirow{2}{*}{BSR6+quark}&100$^4$&1.97&13.13&0.00716\\
&137$^4$&1.99&13.16&0.00684\\
\hline\hline
\end{tabular}
\end{table}

\section{Summary and discussion}\label{four}
In this paper, we introduce the EOS of quark matter with a modification of $2+1$ flavors NJL model to study the structure of hybrid stars. For the gap equation~(\ref{gap}), it's noteworthy that our discussion on the gluon propagator leads to a modification of the coupling constant (the factor of the second term at the right side of Eq.~(\ref{gap})), while the result of Refs.~\cite{PhysRevD.81.116005,OSIPOV200648,OSIPOV20072021} also confirm a modification of that despite in a different form. In addition, for hybrid stars whose temperature is regarded to be zero temperature, the critical baryon chemical potential is generally believed to be 1-1.2 GeV, which is in accordance with the result of our calculation and closer to the regime of a weak 8q coupling in Ref.~\cite{PhysRevD.81.116005}. Therefore, the modification with multi-quark interactions will be relatively small compared with the general $2+1$ flavors NJL model. And our model can be regarded as highlighting the major modification of $G$ (from the feedback of quark propagator to the gluon propagator), then absorbing the small modification of $G$ (from the weak 8q interaction) into the coupling constant $G_1$.

As for the EOS of hadronic matter, we choose five RMF nuclear models to describe for comparison. Inspired by the viewpoint of a crossover DPT, we construct the hybrid EOSs with a 3-window smooth interpolation approach. Beside that we choose the window parameter as 0.4 GeV to study the star configuration with a relatively soft hybrid EOS. Combined with the parameter constraints on the bag constant from the recent binary neutron star (BNS) merger event GW170817, we set $B$ to be $(137$ MeV$)^4$. By a series of calculation, we find the sound velocities of hybrid stars are much smaller than the speed of light, which proves the hybrid EOSs are relatively soft. In the end, we integrate the TOV equations to get the M-R and M-$\rho_c$ relation, which demonstrates that the maximum masses of the hybrid stars do not differ too much and they are all around 2.0 times of solar mass. Furthermore, some previous studies~\cite{PhysRevC.60.025801,PhysRevLett.117.032501} claim that there is no pure quark cores in the center of the hybrid star.  But it is obviously different from the conclusion given by Refs. \cite{PhysRevC.60.025801,PhysRevLett.117.032501}, our present work shows that the heaviest stable stars have central densities higher than that of the deconfinement transition thus suggesting a pure quark core in the hybrid star. And the results are consistent with the mass constraints of the PSR J0348+0432 and PSR J1614-2230.

\acknowledgments

This work is supported in part by the National Natural Science Foundation of China (under Grants No. 11475085, No. 11535005, No. 11690030, and No. 11473012), the Fundamental Research Funds for the Central Universities (under Grant No. 020414380074), the National Basic Research Program of China ("973" Program, Grant No. 2014CB845800), the Strategic Priority Research Program of the Chinese Academy of Sciences
"Multi-waveband Gravitational Wave Universe" (Grant No. XDB23040000) and by the National Major state Basic Research and Development of China (Grant No. 2016YFE0129300).

\section{Appendix}\label{five}
Inspired by the QCD sum rule~\cite{doi:10.1143/PTPS.131.1}, the authors of Ref.~\cite{PhysRevD.85.034031} come up with a relatively simple way to extract the quark's feedback from the gluon propagator. Within the OPE approach, the current-current correlation function can be expressed via a series of local scalar operators' vacuum expectation values, that is, the vacuum condensates. These condensates are regarded as parameters in the QCD sum rules and characterize the nonperturbative property of QCD. And they have also been calculated with LQCD~\cite{PhysRevD.87.034503} and many effective models such as DSEs~\cite{PhysRevC.56.3369,PhysRevD.67.074004}. For the gluon propagator, the quark condensate, which has the lowest dimension among all the condensates, should be contained in the gluon self energy. The vacuum polarization tensor of the gluon involves the term~\cite{Steele1989,PhysRevD.85.034031}
\begin{eqnarray}
  \Pi_{\mu\nu}^{Q} &=& -g^2 \int d^4(y-z)\int\frac{d^4q}{(2\pi)^4}e^{i(p-q)\cdot(y-z)}\nonumber\\
                     \,\,&\times&tr[\gamma_{\mu}\frac{1}{i\slash{q}+m}\gamma_{\nu}\langle\bar{\psi}(y)\psi(z)\rangle]\nonumber\\
                   &=& P_{\mu\nu}(k)k^2\Pi^Q(k^2)\nonumber\\
                   &=& -P_{\mu\nu}(k)\frac{g^2m\langle\bar{\psi}\psi\rangle}{3k^2}+\ldots,\,\,\label{gvpt}
\end{eqnarray}
here $m\langle\bar{\psi}\psi\rangle=m_u\phi_u+m_d\phi_d$, the ellipsis stands for the higher order terms in $m^2/k^2$ (which are neglected), and the superscript $Q$ represents quark. Then, we can extract a quark-unaffected part $D^q$ from the full gluon propagator (where $q$ means quenched). Now the full gluon propagator can be divided in two parts,
\begin{eqnarray}
  D_{\mu\nu}(k) &=& P_{\mu\nu}D(k^2)\nonumber\\
   &=& P_{\mu\nu}(D^q(k^2)+D^Q(k^2)).\,\,\label{fgp}
\end{eqnarray}
And then, considering the DSE of the gluon propagator, we will have
\begin{equation}\label{dsefgp}
  D_{\mu\nu}(k)=D_{\mu\nu}^q(k)+D_{\mu\rho}^0(k)\Pi_{\rho\sigma}^Q(k)D_{\sigma\nu},
\end{equation}
\begin{figure}
\includegraphics[width=0.47\textwidth]{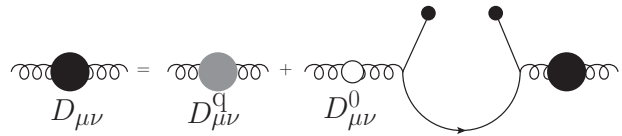}
\caption{The gluon DSE has a vacuum polarization term in which the local quark condensate is contained. See Eqs.~(\ref{gvpt}) and (\ref{dsefgp}).}
\label{Fig:gluonDSE}
\end{figure}
which is shown diagrammatically in Fig.~\ref{Fig:gluonDSE}. With the Eqs.~(\ref{gvpt}),(\ref{fgp}),(\ref{dsefgp}) above, we have
\begin{eqnarray}
  D(k^2) &=& \frac{D^q(k^2)}{1+g^2m\langle\bar{\psi}\psi\rangle_0/ 3k^4}\nonumber\\
   &\approx& \frac{D^q(k^2)}{1+\langle\bar{\psi}\psi\rangle_0/\Lambda^3},\,\,\label{gpz}
\end{eqnarray}
where the subscript refers to zero temperature and zero chemical potential. In addition, just like the Ref.~\cite{PhysRevD.85.034031}, the momentum scale $\Lambda$ is introduced, which is treated as a parameter and absorbs the constants $m$, $g$ as well as the momentum $k$ in our model. With this simplification, the gluon propagator is finite for the infrared region. On the other hand, the ultraviolet region won't be influenced because of the heavily ultraviolet suppression for $D^q(k^2)$.

Then we will extend Eq.~(\ref{gpz}) to the regime of finite temperature and chemical potential by $k\rightarrow k_l=(\overrightarrow{k},\Omega_l)$ and $\langle\bar{\psi}\psi\rangle_0\rightarrow \langle\bar{\psi}\psi\rangle_{T,\mu}$. Combined with the Maclaurin expansion method, we will get the following results,
\begin{widetext}
\begin{eqnarray}
  D(\overrightarrow{k}^2+\Omega_l^2) &=& \frac{D^q(\overrightarrow{k}^2+\Omega_l^2)}{1+\langle\bar{\psi}\psi\rangle_{T,\mu}/\Lambda^3}\nonumber\\
   &=& D^q(\overrightarrow{k}^2+\Omega_l^2)\big|_{\langle\bar{\psi}\psi\rangle_{T,\mu}=0}+\left(D_q^{\prime}(\overrightarrow{k}^2+\Omega_l^2)\big|_{\langle\bar{\psi}\psi\rangle_{T,\mu}=0}-\frac{D^{q}(\overrightarrow{k}^2+\Omega_l^2)\big|_{\langle\bar{\psi}\psi\rangle_{T,\mu}=0}}{\Lambda^3}\right)\langle\bar{\psi}\psi\rangle_{T,\mu}+\ldots\,\label{gpf}
\end{eqnarray}
\end{widetext}
Here $D_q^{\prime}(\overrightarrow{k}^2+\Omega_l^2)$ in the second equation represents taking the derivative of $\langle\bar{\psi}\psi\rangle_{T,\mu}$, and the ellipsis refers to the higher order terms of $\langle\bar{\psi}\psi\rangle_{T,\mu}$, which is generally believed to be less contribution compared with the second term in the second equation. Thus for simplification, it is absorbed in the first term $D^q(\overrightarrow{k}^2+\Omega_l^2)\big|_{\langle\bar{\psi}\psi\rangle_{T,\mu}=0}$ in the following calculation. Additionally, the factor $\left(D_q^{\prime}(\overrightarrow{k}^2+\Omega_l^2)\big|_{\langle\bar{\psi}\psi\rangle_{T,\mu}=0}-\frac{D^{q}(\overrightarrow{k}^2+\Omega_l^2)\big|_{\langle\bar{\psi}\psi\rangle_{T,\mu}=0}}{\Lambda^3}\right)$ is treated as a parameter for simplification in our model. In the normal NJL model, it is equivalent to modifying the coupling constant $G$ as follows:
\begin{equation}\label{cc}
  G=G_1+G_2\langle\bar{\psi}\psi\rangle,
\end{equation}
which is exactly the modification we introduce in this paper.

\bibliography{reference}

\begin{thebibliography}{85}%
\makeatletter
\providecommand \@ifxundefined [1]{%
 \@ifx{#1\undefined}
}%
\providecommand \@ifnum [1]{%
 \ifnum #1\expandafter \@firstoftwo
 \else \expandafter \@secondoftwo
 \fi
}%
\providecommand \@ifx [1]{%
 \ifx #1\expandafter \@firstoftwo
 \else \expandafter \@secondoftwo
 \fi
}%
\providecommand \natexlab [1]{#1}%
\providecommand \enquote  [1]{``#1''}%
\providecommand \bibnamefont  [1]{#1}%
\providecommand \bibfnamefont [1]{#1}%
\providecommand \citenamefont [1]{#1}%
\providecommand \href@noop [0]{\@secondoftwo}%
\providecommand \href [0]{\begingroup \@sanitize@url \@href}%
\providecommand \@href[1]{\@@startlink{#1}\@@href}%
\providecommand \@@href[1]{\endgroup#1\@@endlink}%
\providecommand \@sanitize@url [0]{\catcode `\\12\catcode `\$12\catcode
  `\&12\catcode `\#12\catcode `\^12\catcode `\_12\catcode `\%12\relax}%
\providecommand \@@startlink[1]{}%
\providecommand \@@endlink[0]{}%
\providecommand \url  [0]{\begingroup\@sanitize@url \@url }%
\providecommand \@url [1]{\endgroup\@href {#1}{\urlprefix }}%
\providecommand \urlprefix  [0]{URL }%
\providecommand \Eprint [0]{\href }%
\providecommand \doibase [0]{http://dx.doi.org/}%
\providecommand \selectlanguage [0]{\@gobble}%
\providecommand \bibinfo  [0]{\@secondoftwo}%
\providecommand \bibfield  [0]{\@secondoftwo}%
\providecommand \translation [1]{[#1]}%
\providecommand \BibitemOpen [0]{}%
\providecommand \bibitemStop [0]{}%
\providecommand \bibitemNoStop [0]{.\EOS\space}%
\providecommand \EOS [0]{\spacefactor3000\relax}%
\providecommand \BibitemShut  [1]{\csname bibitem#1\endcsname}%
\let\auto@bib@innerbib\@empty
\bibitem [{\citenamefont {Borsanyi}\ \emph {et~al.}(2010)\citenamefont
  {Borsanyi}, \citenamefont {Fodor}, \citenamefont {Hoelbling}, \citenamefont
  {Katz}, \citenamefont {Krieg}, \citenamefont {Ratti},\ and\ \citenamefont
  {Szabo}}]{Borsanyi:2010bpa}%
  \BibitemOpen
  \bibfield  {author} {\bibinfo {author} {\bibfnamefont {S.}~\bibnamefont
  {Borsanyi}}, \bibinfo {author} {\bibfnamefont {Z.}~\bibnamefont {Fodor}},
  \bibinfo {author} {\bibfnamefont {C.}~\bibnamefont {Hoelbling}}, \bibinfo
  {author} {\bibfnamefont {S.~D.}\ \bibnamefont {Katz}}, \bibinfo {author}
  {\bibfnamefont {S.}~\bibnamefont {Krieg}}, \bibinfo {author} {\bibfnamefont
  {C.}~\bibnamefont {Ratti}}, \ and\ \bibinfo {author} {\bibfnamefont {K.~K.}\
  \bibnamefont {Szabo}} (\bibinfo {collaboration} {Wuppertal-Budapest}),\
  }\href {\doibase 10.1007/JHEP09(2010)073} {\bibfield  {journal} {\bibinfo
  {journal} {JHEP}\ }\textbf {\bibinfo {volume} {09}},\ \bibinfo {pages} {073}
  (\bibinfo {year} {2010})},\ \Eprint {http://arxiv.org/abs/1005.3508}
  {arXiv:1005.3508 [hep-lat]} \BibitemShut {NoStop}%
\bibitem [{\citenamefont {Ejiri}\ and\ \citenamefont
  {Yamada}(2013)}]{PhysRevLett.110.172001}%
  \BibitemOpen
  \bibfield  {author} {\bibinfo {author} {\bibfnamefont {S.}~\bibnamefont
  {Ejiri}}\ and\ \bibinfo {author} {\bibfnamefont {N.}~\bibnamefont {Yamada}},\
  }\href {\doibase 10.1103/PhysRevLett.110.172001} {\bibfield  {journal}
  {\bibinfo  {journal} {Phys. Rev. Lett.}\ }\textbf {\bibinfo {volume} {110}},\
  \bibinfo {pages} {172001} (\bibinfo {year} {2013})}\BibitemShut {NoStop}%
\bibitem [{\citenamefont {Roberts}\ and\ \citenamefont
  {Williams}(1994)}]{ROBERTS1994477}%
  \BibitemOpen
  \bibfield  {author} {\bibinfo {author} {\bibfnamefont {C.~D.}\ \bibnamefont
  {Roberts}}\ and\ \bibinfo {author} {\bibfnamefont {A.~G.}\ \bibnamefont
  {Williams}},\ }\href {\doibase
  http://dx.doi.org/10.1016/0146-6410(94)90049-3} {\bibfield  {journal}
  {\bibinfo  {journal} {Prog. Part. Nucl. Phys.}\ }\textbf {\bibinfo {volume}
  {33}},\ \bibinfo {pages} {477 } (\bibinfo {year} {1994})}\BibitemShut
  {NoStop}%
\bibitem [{\citenamefont {Roberts}\ and\ \citenamefont
  {Schmidt}(2000)}]{Roberts2000S1}%
  \BibitemOpen
  \bibfield  {author} {\bibinfo {author} {\bibfnamefont {C.}~\bibnamefont
  {Roberts}}\ and\ \bibinfo {author} {\bibfnamefont {S.}~\bibnamefont
  {Schmidt}},\ }\href {\doibase
  http://dx.doi.org/10.1016/S0146-6410(00)90011-5} {\bibfield  {journal}
  {\bibinfo  {journal} {Prog. Part. Nucl. Phys.}\ }\textbf {\bibinfo {volume}
  {45, Supplement 1}},\ \bibinfo {pages} {S1 } (\bibinfo {year}
  {2000})}\BibitemShut {NoStop}%
\bibitem [{\citenamefont {Maris}\ and\ \citenamefont
  {Roberts}(2003)}]{doi:10.1142/S0218301303001326}%
  \BibitemOpen
  \bibfield  {author} {\bibinfo {author} {\bibfnamefont {P.}~\bibnamefont
  {Maris}}\ and\ \bibinfo {author} {\bibfnamefont {C.~D.}\ \bibnamefont
  {Roberts}},\ }\href {\doibase 10.1142/S0218301303001326} {\bibfield
  {journal} {\bibinfo  {journal} {Int. J. Mod. Phys. E}\ }\textbf {\bibinfo
  {volume} {12}},\ \bibinfo {pages} {297} (\bibinfo {year} {2003})}\BibitemShut
  {NoStop}%
\bibitem [{\citenamefont {Cl{\"o}et}\ and\ \citenamefont
  {Roberts}(2014)}]{Cloet20141}%
  \BibitemOpen
  \bibfield  {author} {\bibinfo {author} {\bibfnamefont {I.~C.}\ \bibnamefont
  {Cl{\"o}et}}\ and\ \bibinfo {author} {\bibfnamefont {C.~D.}\ \bibnamefont
  {Roberts}},\ }\href {\doibase http://dx.doi.org/10.1016/j.ppnp.2014.02.001}
  {\bibfield  {journal} {\bibinfo  {journal} {Prog. Part. Nucl. Phys.}\
  }\textbf {\bibinfo {volume} {77}},\ \bibinfo {pages} {1 } (\bibinfo {year}
  {2014})}\BibitemShut {NoStop}%
\bibitem [{\citenamefont {Zhao}\ \emph {et~al.}(2014)\citenamefont {Zhao},
  \citenamefont {Cui}, \citenamefont {Jiang},\ and\ \citenamefont
  {Zong}}]{PhysRevD.90.114031}%
  \BibitemOpen
  \bibfield  {author} {\bibinfo {author} {\bibfnamefont {A.-M.}\ \bibnamefont
  {Zhao}}, \bibinfo {author} {\bibfnamefont {Z.-F.}\ \bibnamefont {Cui}},
  \bibinfo {author} {\bibfnamefont {Y.}~\bibnamefont {Jiang}}, \ and\ \bibinfo
  {author} {\bibfnamefont {H.-S.}\ \bibnamefont {Zong}},\ }\href {\doibase
  10.1103/PhysRevD.90.114031} {\bibfield  {journal} {\bibinfo  {journal} {Phys.
  Rev. D}\ }\textbf {\bibinfo {volume} {90}},\ \bibinfo {pages} {114031}
  (\bibinfo {year} {2014})}\BibitemShut {NoStop}%
\bibitem [{\citenamefont {Wang}\ \emph {et~al.}(2015)\citenamefont {Wang},
  \citenamefont {Wang}, \citenamefont {Cui},\ and\ \citenamefont
  {Zong}}]{PhysRevD.91.034017}%
  \BibitemOpen
  \bibfield  {author} {\bibinfo {author} {\bibfnamefont {B.}~\bibnamefont
  {Wang}}, \bibinfo {author} {\bibfnamefont {Y.-L.}\ \bibnamefont {Wang}},
  \bibinfo {author} {\bibfnamefont {Z.-F.}\ \bibnamefont {Cui}}, \ and\
  \bibinfo {author} {\bibfnamefont {H.-S.}\ \bibnamefont {Zong}},\ }\href
  {\doibase 10.1103/PhysRevD.91.034017} {\bibfield  {journal} {\bibinfo
  {journal} {Phys. Rev. D}\ }\textbf {\bibinfo {volume} {91}},\ \bibinfo
  {pages} {034017} (\bibinfo {year} {2015})}\BibitemShut {NoStop}%
\bibitem [{\citenamefont {Xu}\ \emph {et~al.}(2015)\citenamefont {Xu},
  \citenamefont {Cui}, \citenamefont {Wang}, \citenamefont {Shi}, \citenamefont
  {Yang},\ and\ \citenamefont {Zong}}]{PhysRevD.91.056003}%
  \BibitemOpen
  \bibfield  {author} {\bibinfo {author} {\bibfnamefont {S.-S.}\ \bibnamefont
  {Xu}}, \bibinfo {author} {\bibfnamefont {Z.-F.}\ \bibnamefont {Cui}},
  \bibinfo {author} {\bibfnamefont {B.}~\bibnamefont {Wang}}, \bibinfo {author}
  {\bibfnamefont {Y.-M.}\ \bibnamefont {Shi}}, \bibinfo {author} {\bibfnamefont
  {Y.-C.}\ \bibnamefont {Yang}}, \ and\ \bibinfo {author} {\bibfnamefont
  {H.-S.}\ \bibnamefont {Zong}},\ }\href {\doibase 10.1103/PhysRevD.91.056003}
  {\bibfield  {journal} {\bibinfo  {journal} {Phys. Rev. D}\ }\textbf {\bibinfo
  {volume} {91}},\ \bibinfo {pages} {056003} (\bibinfo {year}
  {2015})}\BibitemShut {NoStop}%
\bibitem [{\citenamefont {Pisarski}(1984)}]{PhysRevD.29.2423}%
  \BibitemOpen
  \bibfield  {author} {\bibinfo {author} {\bibfnamefont {R.~D.}\ \bibnamefont
  {Pisarski}},\ }\href {\doibase 10.1103/PhysRevD.29.2423} {\bibfield
  {journal} {\bibinfo  {journal} {Phys. Rev. D}\ }\textbf {\bibinfo {volume}
  {29}},\ \bibinfo {pages} {2423} (\bibinfo {year} {1984})}\BibitemShut
  {NoStop}%
\bibitem [{\citenamefont {Yin}\ \emph {et~al.}(2014)\citenamefont {Yin},
  \citenamefont {Shi}, \citenamefont {Cui}, \citenamefont {Feng},\ and\
  \citenamefont {Zong}}]{PhysRevD.90.036007}%
  \BibitemOpen
  \bibfield  {author} {\bibinfo {author} {\bibfnamefont {P.-L.}\ \bibnamefont
  {Yin}}, \bibinfo {author} {\bibfnamefont {Y.-M.}\ \bibnamefont {Shi}},
  \bibinfo {author} {\bibfnamefont {Z.-F.}\ \bibnamefont {Cui}}, \bibinfo
  {author} {\bibfnamefont {H.-T.}\ \bibnamefont {Feng}}, \ and\ \bibinfo
  {author} {\bibfnamefont {H.-S.}\ \bibnamefont {Zong}},\ }\href {\doibase
  10.1103/PhysRevD.90.036007} {\bibfield  {journal} {\bibinfo  {journal} {Phys.
  Rev. D}\ }\textbf {\bibinfo {volume} {90}},\ \bibinfo {pages} {036007}
  (\bibinfo {year} {2014})}\BibitemShut {NoStop}%
\bibitem [{\citenamefont {Li}\ \emph {et~al.}(2014)\citenamefont {Li},
  \citenamefont {Hou}, \citenamefont {Cui}, \citenamefont {Feng}, \citenamefont
  {Jiang},\ and\ \citenamefont {Zong}}]{PhysRevD.90.073013}%
  \BibitemOpen
  \bibfield  {author} {\bibinfo {author} {\bibfnamefont {J.-F.}\ \bibnamefont
  {Li}}, \bibinfo {author} {\bibfnamefont {F.-Y.}\ \bibnamefont {Hou}},
  \bibinfo {author} {\bibfnamefont {Z.-F.}\ \bibnamefont {Cui}}, \bibinfo
  {author} {\bibfnamefont {H.-T.}\ \bibnamefont {Feng}}, \bibinfo {author}
  {\bibfnamefont {Y.}~\bibnamefont {Jiang}}, \ and\ \bibinfo {author}
  {\bibfnamefont {H.-S.}\ \bibnamefont {Zong}},\ }\href {\doibase
  10.1103/PhysRevD.90.073013} {\bibfield  {journal} {\bibinfo  {journal} {Phys.
  Rev. D}\ }\textbf {\bibinfo {volume} {90}},\ \bibinfo {pages} {073013}
  (\bibinfo {year} {2014})}\BibitemShut {NoStop}%
\bibitem [{\citenamefont {Klevansky}(1992)}]{RevModPhys.64.649}%
  \BibitemOpen
  \bibfield  {author} {\bibinfo {author} {\bibfnamefont {S.~P.}\ \bibnamefont
  {Klevansky}},\ }\href {\doibase 10.1103/RevModPhys.64.649} {\bibfield
  {journal} {\bibinfo  {journal} {Rev. Mod. Phys.}\ }\textbf {\bibinfo {volume}
  {64}},\ \bibinfo {pages} {649} (\bibinfo {year} {1992})}\BibitemShut
  {NoStop}%
\bibitem [{\citenamefont {Buballa}(2005)}]{Buballa2005205}%
  \BibitemOpen
  \bibfield  {author} {\bibinfo {author} {\bibfnamefont {M.}~\bibnamefont
  {Buballa}},\ }\href {\doibase
  http://dx.doi.org/10.1016/j.physrep.2004.11.004} {\bibfield  {journal}
  {\bibinfo  {journal} {Phys. Rep.}\ }\textbf {\bibinfo {volume} {407}},\
  \bibinfo {pages} {205 } (\bibinfo {year} {2005})}\BibitemShut {NoStop}%
\bibitem [{\citenamefont {Cui}\ \emph {et~al.}(2013)\citenamefont {Cui},
  \citenamefont {Shi}, \citenamefont {Xia}, \citenamefont {Jiang},\ and\
  \citenamefont {Zong}}]{Cui2013}%
  \BibitemOpen
  \bibfield  {author} {\bibinfo {author} {\bibfnamefont {Z.-F.}\ \bibnamefont
  {Cui}}, \bibinfo {author} {\bibfnamefont {C.}~\bibnamefont {Shi}}, \bibinfo
  {author} {\bibfnamefont {Y.-H.}\ \bibnamefont {Xia}}, \bibinfo {author}
  {\bibfnamefont {Y.}~\bibnamefont {Jiang}}, \ and\ \bibinfo {author}
  {\bibfnamefont {H.-S.}\ \bibnamefont {Zong}},\ }\href {\doibase
  10.1140/epjc/s10052-013-2612-6} {\bibfield  {journal} {\bibinfo  {journal}
  {Eur. Phys. J. C}\ }\textbf {\bibinfo {volume} {73}},\ \bibinfo {pages}
  {2612} (\bibinfo {year} {2013})}\BibitemShut {NoStop}%
\bibitem [{\citenamefont {Kohyama}\ \emph {et~al.}(2015)\citenamefont
  {Kohyama}, \citenamefont {Kimura},\ and\ \citenamefont
  {Inagaki}}]{NuclPhysB.896.682}%
  \BibitemOpen
  \bibfield  {author} {\bibinfo {author} {\bibfnamefont {H.}~\bibnamefont
  {Kohyama}}, \bibinfo {author} {\bibfnamefont {D.}~\bibnamefont {Kimura}}, \
  and\ \bibinfo {author} {\bibfnamefont {T.}~\bibnamefont {Inagaki}},\
  }\href@noop {} {\bibfield  {journal} {\bibinfo  {journal} {Nucl. Phys. B}\
  }\textbf {\bibinfo {volume} {896}},\ \bibinfo {pages} {682} (\bibinfo {year}
  {2015})}\BibitemShut {NoStop}%
\bibitem [{\citenamefont {Fan}\ \emph {et~al.}(2017{\natexlab{a}})\citenamefont
  {Fan}, \citenamefont {Luo},\ and\ \citenamefont
  {Zong}}]{doi:10.1142/S0217751X17500610}%
  \BibitemOpen
  \bibfield  {author} {\bibinfo {author} {\bibfnamefont {W.}~\bibnamefont
  {Fan}}, \bibinfo {author} {\bibfnamefont {X.}~\bibnamefont {Luo}}, \ and\
  \bibinfo {author} {\bibfnamefont {H.-S.}\ \bibnamefont {Zong}},\ }\href
  {\doibase 10.1142/S0217751X17500610} {\bibfield  {journal} {\bibinfo
  {journal} {Int. J. Mod. Phys. A}\ }\textbf {\bibinfo {volume} {32}},\
  \bibinfo {pages} {1750061} (\bibinfo {year}
  {2017}{\natexlab{a}})}\BibitemShut {NoStop}%
\bibitem [{\citenamefont {Pereira}\ \emph {et~al.}(2016)\citenamefont
  {Pereira}, \citenamefont {Costa},\ and\ \citenamefont
  {Provid\^encia}}]{PhysRevD.94.094001}%
  \BibitemOpen
  \bibfield  {author} {\bibinfo {author} {\bibfnamefont {R.~C.}\ \bibnamefont
  {Pereira}}, \bibinfo {author} {\bibfnamefont {P.}~\bibnamefont {Costa}}, \
  and\ \bibinfo {author} {\bibfnamefont {C.~m.~c.}\ \bibnamefont
  {Provid\^encia}},\ }\href {\doibase 10.1103/PhysRevD.94.094001} {\bibfield
  {journal} {\bibinfo  {journal} {Phys. Rev. D}\ }\textbf {\bibinfo {volume}
  {94}},\ \bibinfo {pages} {094001} (\bibinfo {year} {2016})}\BibitemShut
  {NoStop}%
\bibitem [{\citenamefont {Buballa}\ \emph {et~al.}(2004)\citenamefont
  {Buballa}, \citenamefont {Neumann}, \citenamefont {Oertel},\ and\
  \citenamefont {Shovkovy}}]{BUBALLA200436}%
  \BibitemOpen
  \bibfield  {author} {\bibinfo {author} {\bibfnamefont {M.}~\bibnamefont
  {Buballa}}, \bibinfo {author} {\bibfnamefont {F.}~\bibnamefont {Neumann}},
  \bibinfo {author} {\bibfnamefont {M.}~\bibnamefont {Oertel}}, \ and\ \bibinfo
  {author} {\bibfnamefont {I.}~\bibnamefont {Shovkovy}},\ }\href {\doibase
  https://doi.org/10.1016/j.physletb.2004.05.064} {\bibfield  {journal}
  {\bibinfo  {journal} {Phys. Lett. B}\ }\textbf {\bibinfo {volume} {595}},\
  \bibinfo {pages} {36 } (\bibinfo {year} {2004})}\BibitemShut {NoStop}%
\bibitem [{\citenamefont {Kl{\"a}hn}\ \emph {et~al.}(2007)\citenamefont
  {Kl{\"a}hn}, \citenamefont {Blaschke}, \citenamefont {Sandin}, \citenamefont
  {Fuchs}, \citenamefont {Faessler}, \citenamefont {Grigorian}, \citenamefont
  {R{\"o}pke},\ and\ \citenamefont {Tr{\"u}mper}}]{KLAHN2007170}%
  \BibitemOpen
  \bibfield  {author} {\bibinfo {author} {\bibfnamefont {T.}~\bibnamefont
  {Kl{\"a}hn}}, \bibinfo {author} {\bibfnamefont {D.}~\bibnamefont {Blaschke}},
  \bibinfo {author} {\bibfnamefont {F.}~\bibnamefont {Sandin}}, \bibinfo
  {author} {\bibfnamefont {C.}~\bibnamefont {Fuchs}}, \bibinfo {author}
  {\bibfnamefont {A.}~\bibnamefont {Faessler}}, \bibinfo {author}
  {\bibfnamefont {H.}~\bibnamefont {Grigorian}}, \bibinfo {author}
  {\bibfnamefont {G.}~\bibnamefont {R{\"o}pke}}, \ and\ \bibinfo {author}
  {\bibfnamefont {J.}~\bibnamefont {Tr{\"u}mper}},\ }\href {\doibase
  https://doi.org/10.1016/j.physletb.2007.08.048} {\bibfield  {journal}
  {\bibinfo  {journal} {Phys. Lett. B}\ }\textbf {\bibinfo {volume} {654}},\
  \bibinfo {pages} {170 } (\bibinfo {year} {2007})}\BibitemShut {NoStop}%
\bibitem [{\citenamefont {Fan}\ \emph {et~al.}(2016)\citenamefont {Fan},
  \citenamefont {Luo},\ and\ \citenamefont {Zong}}]{2016arXiv160807903F}%
  \BibitemOpen
  \bibfield  {author} {\bibinfo {author} {\bibfnamefont {W.}~\bibnamefont
  {Fan}}, \bibinfo {author} {\bibfnamefont {X.}~\bibnamefont {Luo}}, \ and\
  \bibinfo {author} {\bibfnamefont {H.-S.}\ \bibnamefont {Zong}},\ }\href@noop
  {} {\bibfield  {journal} {\bibinfo  {journal} {ArXiv e-prints}\ } (\bibinfo
  {year} {2016})},\ \Eprint {http://arxiv.org/abs/1608.07903} {arXiv:1608.07903
  [hep-ph]} \BibitemShut {NoStop}%
\bibitem [{\citenamefont {Steele}(1989)}]{Steele1989}%
  \BibitemOpen
  \bibfield  {author} {\bibinfo {author} {\bibfnamefont {T.~G.}\ \bibnamefont
  {Steele}},\ }\href {\doibase 10.1007/BF01548457} {\bibfield  {journal}
  {\bibinfo  {journal} {Z. Phys. C}\ }\textbf {\bibinfo {volume} {42}},\
  \bibinfo {pages} {499} (\bibinfo {year} {1989})}\BibitemShut {NoStop}%
\bibitem [{\citenamefont {Jiang}\ \emph {et~al.}(2012)\citenamefont {Jiang},
  \citenamefont {Gong}, \citenamefont {Sun},\ and\ \citenamefont
  {Zong}}]{PhysRevD.85.034031}%
  \BibitemOpen
  \bibfield  {author} {\bibinfo {author} {\bibfnamefont {Y.}~\bibnamefont
  {Jiang}}, \bibinfo {author} {\bibfnamefont {H.}~\bibnamefont {Gong}},
  \bibinfo {author} {\bibfnamefont {W.-M.}\ \bibnamefont {Sun}}, \ and\
  \bibinfo {author} {\bibfnamefont {H.-S.}\ \bibnamefont {Zong}},\ }\href
  {\doibase 10.1103/PhysRevD.85.034031} {\bibfield  {journal} {\bibinfo
  {journal} {Phys. Rev. D}\ }\textbf {\bibinfo {volume} {85}},\ \bibinfo
  {pages} {034031} (\bibinfo {year} {2012})}\BibitemShut {NoStop}%
\bibitem [{\citenamefont {Shi}\ \emph {et~al.}(2016)\citenamefont {Shi},
  \citenamefont {Du}, \citenamefont {Xu}, \citenamefont {Liu},\ and\
  \citenamefont {Zong}}]{PhysRevD.93.036006}%
  \BibitemOpen
  \bibfield  {author} {\bibinfo {author} {\bibfnamefont {C.}~\bibnamefont
  {Shi}}, \bibinfo {author} {\bibfnamefont {Y.-L.}\ \bibnamefont {Du}},
  \bibinfo {author} {\bibfnamefont {S.-S.}\ \bibnamefont {Xu}}, \bibinfo
  {author} {\bibfnamefont {X.-J.}\ \bibnamefont {Liu}}, \ and\ \bibinfo
  {author} {\bibfnamefont {H.-S.}\ \bibnamefont {Zong}},\ }\href {\doibase
  10.1103/PhysRevD.93.036006} {\bibfield  {journal} {\bibinfo  {journal} {Phys.
  Rev. D}\ }\textbf {\bibinfo {volume} {93}},\ \bibinfo {pages} {036006}
  (\bibinfo {year} {2016})}\BibitemShut {NoStop}%
\bibitem [{\citenamefont {Ratti}\ \emph {et~al.}(2007)\citenamefont {Ratti},
  \citenamefont {R{\"o}{\ss}ner}, \citenamefont {Thaler},\ and\ \citenamefont
  {Weise}}]{Ratti2007}%
  \BibitemOpen
  \bibfield  {author} {\bibinfo {author} {\bibfnamefont {C.}~\bibnamefont
  {Ratti}}, \bibinfo {author} {\bibfnamefont {S.}~\bibnamefont
  {R{\"o}{\ss}ner}}, \bibinfo {author} {\bibfnamefont {M.}~\bibnamefont
  {Thaler}}, \ and\ \bibinfo {author} {\bibfnamefont {W.}~\bibnamefont
  {Weise}},\ }\href {\doibase 10.1140/epjc/s10052-006-0065-x} {\bibfield
  {journal} {\bibinfo  {journal} {Eur. Phys. J. C}\ }\textbf {\bibinfo {volume}
  {49}},\ \bibinfo {pages} {213} (\bibinfo {year} {2007})}\BibitemShut
  {NoStop}%
\bibitem [{\citenamefont {Fukushima}(2008)}]{PhysRevD.77.114028}%
  \BibitemOpen
  \bibfield  {author} {\bibinfo {author} {\bibfnamefont {K.}~\bibnamefont
  {Fukushima}},\ }\href {\doibase 10.1103/PhysRevD.77.114028} {\bibfield
  {journal} {\bibinfo  {journal} {Phys. Rev. D}\ }\textbf {\bibinfo {volume}
  {77}},\ \bibinfo {pages} {114028} (\bibinfo {year} {2008})}\BibitemShut
  {NoStop}%
\bibitem [{\citenamefont {Cui}\ \emph {et~al.}(2014)\citenamefont {Cui},
  \citenamefont {Shi}, \citenamefont {Sun}, \citenamefont {Wang},\ and\
  \citenamefont {Zong}}]{Cui2014}%
  \BibitemOpen
  \bibfield  {author} {\bibinfo {author} {\bibfnamefont {Z.-F.}\ \bibnamefont
  {Cui}}, \bibinfo {author} {\bibfnamefont {C.}~\bibnamefont {Shi}}, \bibinfo
  {author} {\bibfnamefont {W.-M.}\ \bibnamefont {Sun}}, \bibinfo {author}
  {\bibfnamefont {Y.-L.}\ \bibnamefont {Wang}}, \ and\ \bibinfo {author}
  {\bibfnamefont {H.-S.}\ \bibnamefont {Zong}},\ }\href {\doibase
  10.1140/epjc/s10052-014-2782-x} {\bibfield  {journal} {\bibinfo  {journal}
  {Eur. Phys. J. C}\ }\textbf {\bibinfo {volume} {74}},\ \bibinfo {pages}
  {2782} (\bibinfo {year} {2014})}\BibitemShut {NoStop}%
\bibitem [{\citenamefont {Shao}\ \emph {et~al.}(2016)\citenamefont {Shao},
  \citenamefont {Tang}, \citenamefont {Di~Toro}, \citenamefont {Colonna},
  \citenamefont {Gao},\ and\ \citenamefont {Gao}}]{PhysRevD.94.014008}%
  \BibitemOpen
  \bibfield  {author} {\bibinfo {author} {\bibfnamefont {G.-Y.}\ \bibnamefont
  {Shao}}, \bibinfo {author} {\bibfnamefont {Z.-D.}\ \bibnamefont {Tang}},
  \bibinfo {author} {\bibfnamefont {M.}~\bibnamefont {Di~Toro}}, \bibinfo
  {author} {\bibfnamefont {M.}~\bibnamefont {Colonna}}, \bibinfo {author}
  {\bibfnamefont {X.-Y.}\ \bibnamefont {Gao}}, \ and\ \bibinfo {author}
  {\bibfnamefont {N.}~\bibnamefont {Gao}},\ }\href {\doibase
  10.1103/PhysRevD.94.014008} {\bibfield  {journal} {\bibinfo  {journal} {Phys.
  Rev. D}\ }\textbf {\bibinfo {volume} {94}},\ \bibinfo {pages} {014008}
  (\bibinfo {year} {2016})}\BibitemShut {NoStop}%
\bibitem [{\citenamefont {Li}\ \emph {et~al.}(2015)\citenamefont {Li},
  \citenamefont {Zuo},\ and\ \citenamefont {Peng}}]{PhysRevC.91.035803}%
  \BibitemOpen
  \bibfield  {author} {\bibinfo {author} {\bibfnamefont {A.}~\bibnamefont
  {Li}}, \bibinfo {author} {\bibfnamefont {W.}~\bibnamefont {Zuo}}, \ and\
  \bibinfo {author} {\bibfnamefont {G.~X.}\ \bibnamefont {Peng}},\ }\href
  {\doibase 10.1103/PhysRevC.91.035803} {\bibfield  {journal} {\bibinfo
  {journal} {Phys. Rev. C}\ }\textbf {\bibinfo {volume} {91}},\ \bibinfo
  {pages} {035803} (\bibinfo {year} {2015})}\BibitemShut {NoStop}%
\bibitem [{\citenamefont {Mallick}\ and\ \citenamefont
  {Sahu}(2014)}]{Mallick201496}%
  \BibitemOpen
  \bibfield  {author} {\bibinfo {author} {\bibfnamefont {R.}~\bibnamefont
  {Mallick}}\ and\ \bibinfo {author} {\bibfnamefont {P.}~\bibnamefont {Sahu}},\
  }\href {\doibase http://dx.doi.org/10.1016/j.nuclphysa.2013.11.009}
  {\bibfield  {journal} {\bibinfo  {journal} {Nucl. Phys. A}\ }\textbf
  {\bibinfo {volume} {921}},\ \bibinfo {pages} {96 } (\bibinfo {year}
  {2014})}\BibitemShut {NoStop}%
\bibitem [{\citenamefont {Zhao}\ \emph
  {et~al.}(2015{\natexlab{a}})\citenamefont {Zhao}, \citenamefont {Yan},
  \citenamefont {Luo},\ and\ \citenamefont {Zong}}]{PhysRevD.91.034018}%
  \BibitemOpen
  \bibfield  {author} {\bibinfo {author} {\bibfnamefont {T.}~\bibnamefont
  {Zhao}}, \bibinfo {author} {\bibfnamefont {Y.}~\bibnamefont {Yan}}, \bibinfo
  {author} {\bibfnamefont {X.-L.}\ \bibnamefont {Luo}}, \ and\ \bibinfo
  {author} {\bibfnamefont {H.-S.}\ \bibnamefont {Zong}},\ }\href {\doibase
  10.1103/PhysRevD.91.034018} {\bibfield  {journal} {\bibinfo  {journal} {Phys.
  Rev. D}\ }\textbf {\bibinfo {volume} {91}},\ \bibinfo {pages} {034018}
  (\bibinfo {year} {2015}{\natexlab{a}})}\BibitemShut {NoStop}%
\bibitem [{\citenamefont {Benic}\ \emph {et~al.}(2015)\citenamefont {Benic},
  \citenamefont {Blaschke}, \citenamefont {Alvarez-Castillo}, \citenamefont
  {Fischer},\ and\ \citenamefont {Typel}}]{refId0}%
  \BibitemOpen
  \bibfield  {author} {\bibinfo {author} {\bibfnamefont {S.}~\bibnamefont
  {Benic}}, \bibinfo {author} {\bibfnamefont {D.}~\bibnamefont {Blaschke}},
  \bibinfo {author} {\bibfnamefont {D.~E.}\ \bibnamefont {Alvarez-Castillo}},
  \bibinfo {author} {\bibfnamefont {T.}~\bibnamefont {Fischer}}, \ and\
  \bibinfo {author} {\bibfnamefont {S.}~\bibnamefont {Typel}},\ }\href
  {\doibase 10.1051/0004-6361/201425318} {\bibfield  {journal} {\bibinfo
  {journal} {A\&A}\ }\textbf {\bibinfo {volume} {577}},\ \bibinfo {pages} {A40}
  (\bibinfo {year} {2015})}\BibitemShut {NoStop}%
\bibitem [{\citenamefont {Alvarez-Castillo}\ and\ \citenamefont
  {Blaschke}(2015)}]{Alvarez-Castillo2015}%
  \BibitemOpen
  \bibfield  {author} {\bibinfo {author} {\bibfnamefont {D.~E.}\ \bibnamefont
  {Alvarez-Castillo}}\ and\ \bibinfo {author} {\bibfnamefont {D.}~\bibnamefont
  {Blaschke}},\ }\href {\doibase 10.1134/S1063779615050032} {\bibfield
  {journal} {\bibinfo  {journal} {Phys. Part. Nucl.}\ }\textbf {\bibinfo
  {volume} {46}},\ \bibinfo {pages} {846} (\bibinfo {year} {2015})}\BibitemShut
  {NoStop}%
\bibitem [{\citenamefont {de~Forcrand}\ \emph {et~al.}(2014)\citenamefont
  {de~Forcrand}, \citenamefont {Langelage}, \citenamefont {Philipsen},\ and\
  \citenamefont {Unger}}]{PhysRevLett.113.152002}%
  \BibitemOpen
  \bibfield  {author} {\bibinfo {author} {\bibfnamefont {P.}~\bibnamefont
  {de~Forcrand}}, \bibinfo {author} {\bibfnamefont {J.}~\bibnamefont
  {Langelage}}, \bibinfo {author} {\bibfnamefont {O.}~\bibnamefont
  {Philipsen}}, \ and\ \bibinfo {author} {\bibfnamefont {W.}~\bibnamefont
  {Unger}},\ }\href {\doibase 10.1103/PhysRevLett.113.152002} {\bibfield
  {journal} {\bibinfo  {journal} {Phys. Rev. Lett.}\ }\textbf {\bibinfo
  {volume} {113}},\ \bibinfo {pages} {152002} (\bibinfo {year}
  {2014})}\BibitemShut {NoStop}%
\bibitem [{\citenamefont {Endrodi}(2015)}]{Endrodi:2015oba}%
  \BibitemOpen
  \bibfield  {author} {\bibinfo {author} {\bibfnamefont {G.}~\bibnamefont
  {Endrodi}},\ }\href {\doibase 10.1007/JHEP07(2015)173} {\bibfield  {journal}
  {\bibinfo  {journal} {JHEP}\ }\textbf {\bibinfo {volume} {07}},\ \bibinfo
  {pages} {173} (\bibinfo {year} {2015})},\ \Eprint
  {http://arxiv.org/abs/1504.08280} {arXiv:1504.08280 [hep-lat]} \BibitemShut
  {NoStop}%
\bibitem [{\citenamefont {Braguta}\ \emph {et~al.}(2015)\citenamefont
  {Braguta}, \citenamefont {Goy}, \citenamefont {Ilgenfritz}, \citenamefont
  {Kotov}, \citenamefont {Molochkov}, \citenamefont {Muller-Preussker},\ and\
  \citenamefont {Petersson}}]{Braguta:2015zta}%
  \BibitemOpen
  \bibfield  {author} {\bibinfo {author} {\bibfnamefont {V.~V.}\ \bibnamefont
  {Braguta}}, \bibinfo {author} {\bibfnamefont {V.~A.}\ \bibnamefont {Goy}},
  \bibinfo {author} {\bibfnamefont {E.~M.}\ \bibnamefont {Ilgenfritz}},
  \bibinfo {author} {\bibfnamefont {A.~{\relax Yu}.}\ \bibnamefont {Kotov}},
  \bibinfo {author} {\bibfnamefont {A.~V.}\ \bibnamefont {Molochkov}}, \bibinfo
  {author} {\bibfnamefont {M.}~\bibnamefont {Muller-Preussker}}, \ and\
  \bibinfo {author} {\bibfnamefont {B.}~\bibnamefont {Petersson}},\ }\href
  {\doibase 10.1007/JHEP06(2015)094} {\bibfield  {journal} {\bibinfo  {journal}
  {JHEP}\ }\textbf {\bibinfo {volume} {06}},\ \bibinfo {pages} {094} (\bibinfo
  {year} {2015})},\ \Eprint {http://arxiv.org/abs/1503.06670} {arXiv:1503.06670
  [hep-lat]} \BibitemShut {NoStop}%
\bibitem [{\citenamefont {Masuda}\ \emph
  {et~al.}(2013{\natexlab{a}})\citenamefont {Masuda}, \citenamefont {Hatsuda},\
  and\ \citenamefont {Takatsuka}}]{Masuda:2012ed}%
  \BibitemOpen
  \bibfield  {author} {\bibinfo {author} {\bibfnamefont {K.}~\bibnamefont
  {Masuda}}, \bibinfo {author} {\bibfnamefont {T.}~\bibnamefont {Hatsuda}}, \
  and\ \bibinfo {author} {\bibfnamefont {T.}~\bibnamefont {Takatsuka}},\ }\href
  {\doibase 10.1093/ptep/ptt045} {\bibfield  {journal} {\bibinfo  {journal}
  {PTEP}\ }\textbf {\bibinfo {volume} {2013}},\ \bibinfo {pages} {073D01}
  (\bibinfo {year} {2013}{\natexlab{a}})},\ \Eprint
  {http://arxiv.org/abs/1212.6803} {arXiv:1212.6803 [nucl-th]} \BibitemShut
  {NoStop}%
\bibitem [{\citenamefont {Masuda}\ \emph
  {et~al.}(2013{\natexlab{b}})\citenamefont {Masuda}, \citenamefont {Hatsuda},\
  and\ \citenamefont {Takatsuka}}]{0004-637X-764-1-12}%
  \BibitemOpen
  \bibfield  {author} {\bibinfo {author} {\bibfnamefont {K.}~\bibnamefont
  {Masuda}}, \bibinfo {author} {\bibfnamefont {T.}~\bibnamefont {Hatsuda}}, \
  and\ \bibinfo {author} {\bibfnamefont {T.}~\bibnamefont {Takatsuka}},\ }\href
  {http://stacks.iop.org/0004-637X/764/i=1/a=12} {\bibfield  {journal}
  {\bibinfo  {journal} {Astrophys. J}\ }\textbf {\bibinfo {volume} {764}},\
  \bibinfo {pages} {12} (\bibinfo {year} {2013}{\natexlab{b}})}\BibitemShut
  {NoStop}%
\bibitem [{\citenamefont {Kojo}\ \emph {et~al.}(2015)\citenamefont {Kojo},
  \citenamefont {Powell}, \citenamefont {Song},\ and\ \citenamefont
  {Baym}}]{PhysRevD.91.045003}%
  \BibitemOpen
  \bibfield  {author} {\bibinfo {author} {\bibfnamefont {T.}~\bibnamefont
  {Kojo}}, \bibinfo {author} {\bibfnamefont {P.~D.}\ \bibnamefont {Powell}},
  \bibinfo {author} {\bibfnamefont {Y.}~\bibnamefont {Song}}, \ and\ \bibinfo
  {author} {\bibfnamefont {G.}~\bibnamefont {Baym}},\ }\href {\doibase
  10.1103/PhysRevD.91.045003} {\bibfield  {journal} {\bibinfo  {journal} {Phys.
  Rev. D}\ }\textbf {\bibinfo {volume} {91}},\ \bibinfo {pages} {045003}
  (\bibinfo {year} {2015})}\BibitemShut {NoStop}%
\bibitem [{\citenamefont {Zhao}\ \emph {et~al.}(2017)\citenamefont {Zhao},
  \citenamefont {Li}, \citenamefont {Zhao}, \citenamefont {Yan}, \citenamefont
  {Luo},\ and\ \citenamefont {Zong}}]{doi:10.1142/S0217732317500511}%
  \BibitemOpen
  \bibfield  {author} {\bibinfo {author} {\bibfnamefont {T.}~\bibnamefont
  {Zhao}}, \bibinfo {author} {\bibfnamefont {C.-M.}\ \bibnamefont {Li}},
  \bibinfo {author} {\bibfnamefont {Y.-P.}\ \bibnamefont {Zhao}}, \bibinfo
  {author} {\bibfnamefont {Y.}~\bibnamefont {Yan}}, \bibinfo {author}
  {\bibfnamefont {X.-L.}\ \bibnamefont {Luo}}, \ and\ \bibinfo {author}
  {\bibfnamefont {H.-S.}\ \bibnamefont {Zong}},\ }\href {\doibase
  10.1142/S0217732317500511} {\bibfield  {journal} {\bibinfo  {journal} {Mod.
  Phys. Lett. A}\ }\textbf {\bibinfo {volume} {32}},\ \bibinfo {pages}
  {1750051} (\bibinfo {year} {2017})}\BibitemShut {NoStop}%
\bibitem [{\citenamefont {Li}\ \emph {et~al.}(2017)\citenamefont {Li},
  \citenamefont {Zhang}, \citenamefont {Zhao}, \citenamefont {Zhao},\ and\
  \citenamefont {Zong}}]{PhysRevD.95.056018}%
  \BibitemOpen
  \bibfield  {author} {\bibinfo {author} {\bibfnamefont {C.-M.}\ \bibnamefont
  {Li}}, \bibinfo {author} {\bibfnamefont {J.-L.}\ \bibnamefont {Zhang}},
  \bibinfo {author} {\bibfnamefont {T.}~\bibnamefont {Zhao}}, \bibinfo {author}
  {\bibfnamefont {Y.-P.}\ \bibnamefont {Zhao}}, \ and\ \bibinfo {author}
  {\bibfnamefont {H.-S.}\ \bibnamefont {Zong}},\ }\href {\doibase
  10.1103/PhysRevD.95.056018} {\bibfield  {journal} {\bibinfo  {journal} {Phys.
  Rev. D}\ }\textbf {\bibinfo {volume} {95}},\ \bibinfo {pages} {056018}
  (\bibinfo {year} {2017})}\BibitemShut {NoStop}%
\bibitem [{\citenamefont {Zhao}\ \emph
  {et~al.}(2015{\natexlab{b}})\citenamefont {Zhao}, \citenamefont {Xu},
  \citenamefont {Yan}, \citenamefont {Luo}, \citenamefont {Liu},\ and\
  \citenamefont {Zong}}]{PhysRevD.92.054012}%
  \BibitemOpen
  \bibfield  {author} {\bibinfo {author} {\bibfnamefont {T.}~\bibnamefont
  {Zhao}}, \bibinfo {author} {\bibfnamefont {S.-S.}\ \bibnamefont {Xu}},
  \bibinfo {author} {\bibfnamefont {Y.}~\bibnamefont {Yan}}, \bibinfo {author}
  {\bibfnamefont {X.-L.}\ \bibnamefont {Luo}}, \bibinfo {author} {\bibfnamefont
  {X.-J.}\ \bibnamefont {Liu}}, \ and\ \bibinfo {author} {\bibfnamefont
  {H.-S.}\ \bibnamefont {Zong}},\ }\href {\doibase 10.1103/PhysRevD.92.054012}
  {\bibfield  {journal} {\bibinfo  {journal} {Phys. Rev. D}\ }\textbf {\bibinfo
  {volume} {92}},\ \bibinfo {pages} {054012} (\bibinfo {year}
  {2015}{\natexlab{b}})}\BibitemShut {NoStop}%
\bibitem [{\citenamefont {Fan}\ \emph {et~al.}(2017{\natexlab{b}})\citenamefont
  {Fan}, \citenamefont {Fan}, \citenamefont {Wang},\ and\ \citenamefont
  {Zong}}]{doi:10.1142/S0217732317501073}%
  \BibitemOpen
  \bibfield  {author} {\bibinfo {author} {\bibfnamefont {Z.-Y.}\ \bibnamefont
  {Fan}}, \bibinfo {author} {\bibfnamefont {W.-K.}\ \bibnamefont {Fan}},
  \bibinfo {author} {\bibfnamefont {Q.-W.}\ \bibnamefont {Wang}}, \ and\
  \bibinfo {author} {\bibfnamefont {H.-S.}\ \bibnamefont {Zong}},\ }\href
  {\doibase 10.1142/S0217732317501073} {\bibfield  {journal} {\bibinfo
  {journal} {Mod. Phys. Lett. A}\ }\textbf {\bibinfo {volume} {32}},\ \bibinfo
  {pages} {1750107} (\bibinfo {year} {2017}{\natexlab{b}})}\BibitemShut
  {NoStop}%
\bibitem [{\citenamefont {Antoniadis}\ \emph {et~al.}(2013)\citenamefont
  {Antoniadis}, \citenamefont {Freire}, \citenamefont {Wex}, \citenamefont
  {Tauris}, \citenamefont {Lynch}, \citenamefont {van Kerkwijk}, \citenamefont
  {Kramer}, \citenamefont {Bassa}, \citenamefont {Dhillon}, \citenamefont
  {Driebe}, \citenamefont {Hessels}, \citenamefont {Kaspi}, \citenamefont
  {Kondratiev}, \citenamefont {Langer}, \citenamefont {Marsh}, \citenamefont
  {McLaughlin}, \citenamefont {Pennucci}, \citenamefont {Ransom}, \citenamefont
  {Stairs}, \citenamefont {van Leeuwen}, \citenamefont {Verbiest},\ and\
  \citenamefont {Whelan}}]{Antoniadis1233232}%
  \BibitemOpen
  \bibfield  {author} {\bibinfo {author} {\bibfnamefont {J.}~\bibnamefont
  {Antoniadis}}, \bibinfo {author} {\bibfnamefont {P.~C.~C.}\ \bibnamefont
  {Freire}}, \bibinfo {author} {\bibfnamefont {N.}~\bibnamefont {Wex}},
  \bibinfo {author} {\bibfnamefont {T.~M.}\ \bibnamefont {Tauris}}, \bibinfo
  {author} {\bibfnamefont {R.~S.}\ \bibnamefont {Lynch}}, \bibinfo {author}
  {\bibfnamefont {M.~H.}\ \bibnamefont {van Kerkwijk}}, \bibinfo {author}
  {\bibfnamefont {M.}~\bibnamefont {Kramer}}, \bibinfo {author} {\bibfnamefont
  {C.}~\bibnamefont {Bassa}}, \bibinfo {author} {\bibfnamefont {V.~S.}\
  \bibnamefont {Dhillon}}, \bibinfo {author} {\bibfnamefont {T.}~\bibnamefont
  {Driebe}}, \bibinfo {author} {\bibfnamefont {J.~W.~T.}\ \bibnamefont
  {Hessels}}, \bibinfo {author} {\bibfnamefont {V.~M.}\ \bibnamefont {Kaspi}},
  \bibinfo {author} {\bibfnamefont {V.~I.}\ \bibnamefont {Kondratiev}},
  \bibinfo {author} {\bibfnamefont {N.}~\bibnamefont {Langer}}, \bibinfo
  {author} {\bibfnamefont {T.~R.}\ \bibnamefont {Marsh}}, \bibinfo {author}
  {\bibfnamefont {M.~A.}\ \bibnamefont {McLaughlin}}, \bibinfo {author}
  {\bibfnamefont {T.~T.}\ \bibnamefont {Pennucci}}, \bibinfo {author}
  {\bibfnamefont {S.~M.}\ \bibnamefont {Ransom}}, \bibinfo {author}
  {\bibfnamefont {I.~H.}\ \bibnamefont {Stairs}}, \bibinfo {author}
  {\bibfnamefont {J.}~\bibnamefont {van Leeuwen}}, \bibinfo {author}
  {\bibfnamefont {J.~P.~W.}\ \bibnamefont {Verbiest}}, \ and\ \bibinfo {author}
  {\bibfnamefont {D.~G.}\ \bibnamefont {Whelan}},\ }\href@noop {} {\bibfield
  {journal} {\bibinfo  {journal} {Science}\ }\textbf {\bibinfo {volume} {340}}
  (\bibinfo {year} {2013})}\BibitemShut {NoStop}%
\bibitem [{\citenamefont {Fonseca}\ \emph {et~al.}(2016)\citenamefont {Fonseca}
  \emph {et~al.}}]{Fonseca:2016tux}%
  \BibitemOpen
  \bibfield  {author} {\bibinfo {author} {\bibfnamefont {E.}~\bibnamefont
  {Fonseca}} \emph {et~al.},\ }\href {\doibase 10.3847/0004-637X/832/2/167}
  {\bibfield  {journal} {\bibinfo  {journal} {Astrophys. J.}\ }\textbf
  {\bibinfo {volume} {832}},\ \bibinfo {pages} {167} (\bibinfo {year}
  {2016})},\ \Eprint {http://arxiv.org/abs/1603.00545} {arXiv:1603.00545
  [astro-ph.HE]} \BibitemShut {NoStop}%
\bibitem [{\citenamefont {Lalazissis}\ \emph {et~al.}(1997)\citenamefont
  {Lalazissis}, \citenamefont {K\"onig},\ and\ \citenamefont
  {Ring}}]{PhysRevC.55.540}%
  \BibitemOpen
  \bibfield  {author} {\bibinfo {author} {\bibfnamefont {G.~A.}\ \bibnamefont
  {Lalazissis}}, \bibinfo {author} {\bibfnamefont {J.}~\bibnamefont {K\"onig}},
  \ and\ \bibinfo {author} {\bibfnamefont {P.}~\bibnamefont {Ring}},\ }\href
  {\doibase 10.1103/PhysRevC.55.540} {\bibfield  {journal} {\bibinfo  {journal}
  {Phys. Rev. C}\ }\textbf {\bibinfo {volume} {55}},\ \bibinfo {pages} {540}
  (\bibinfo {year} {1997})}\BibitemShut {NoStop}%
\bibitem [{\citenamefont {Horowitz}\ and\ \citenamefont
  {Piekarewicz}(2001)}]{PhysRevLett.86.5647}%
  \BibitemOpen
  \bibfield  {author} {\bibinfo {author} {\bibfnamefont {C.~J.}\ \bibnamefont
  {Horowitz}}\ and\ \bibinfo {author} {\bibfnamefont {J.}~\bibnamefont
  {Piekarewicz}},\ }\href {\doibase 10.1103/PhysRevLett.86.5647} {\bibfield
  {journal} {\bibinfo  {journal} {Phys. Rev. Lett.}\ }\textbf {\bibinfo
  {volume} {86}},\ \bibinfo {pages} {5647} (\bibinfo {year}
  {2001})}\BibitemShut {NoStop}%
\bibitem [{\citenamefont {Banik}\ \emph {et~al.}(2014)\citenamefont {Banik},
  \citenamefont {Hempel},\ and\ \citenamefont
  {Bandyopadhyay}}]{0067-0049-214-2-22}%
  \BibitemOpen
  \bibfield  {author} {\bibinfo {author} {\bibfnamefont {S.}~\bibnamefont
  {Banik}}, \bibinfo {author} {\bibfnamefont {M.}~\bibnamefont {Hempel}}, \
  and\ \bibinfo {author} {\bibfnamefont {D.}~\bibnamefont {Bandyopadhyay}},\
  }\href {http://stacks.iop.org/0067-0049/214/i=2/a=22} {\bibfield  {journal}
  {\bibinfo  {journal} {The Astrophys. J. Suppl. Ser.}\ }\textbf {\bibinfo
  {volume} {214}},\ \bibinfo {pages} {22} (\bibinfo {year} {2014})}\BibitemShut
  {NoStop}%
\bibitem [{\citenamefont {Lalazissis}\ \emph {et~al.}(2005)\citenamefont
  {Lalazissis}, \citenamefont {Nik\ifmmode \check{s}\else
  \v{s}\fi{}i\ifmmode~\acute{c}\else \'{c}\fi{}}, \citenamefont {Vretenar},\
  and\ \citenamefont {Ring}}]{PhysRevC.71.024312}%
  \BibitemOpen
  \bibfield  {author} {\bibinfo {author} {\bibfnamefont {G.~A.}\ \bibnamefont
  {Lalazissis}}, \bibinfo {author} {\bibfnamefont {T.}~\bibnamefont
  {Nik\ifmmode \check{s}\else \v{s}\fi{}i\ifmmode~\acute{c}\else \'{c}\fi{}}},
  \bibinfo {author} {\bibfnamefont {D.}~\bibnamefont {Vretenar}}, \ and\
  \bibinfo {author} {\bibfnamefont {P.}~\bibnamefont {Ring}},\ }\href {\doibase
  10.1103/PhysRevC.71.024312} {\bibfield  {journal} {\bibinfo  {journal} {Phys.
  Rev. C}\ }\textbf {\bibinfo {volume} {71}},\ \bibinfo {pages} {024312}
  (\bibinfo {year} {2005})}\BibitemShut {NoStop}%
\bibitem [{\citenamefont {Dhiman}\ \emph {et~al.}(2007)\citenamefont {Dhiman},
  \citenamefont {Kumar},\ and\ \citenamefont {Agrawal}}]{PhysRevC.76.045801}%
  \BibitemOpen
  \bibfield  {author} {\bibinfo {author} {\bibfnamefont {S.~K.}\ \bibnamefont
  {Dhiman}}, \bibinfo {author} {\bibfnamefont {R.}~\bibnamefont {Kumar}}, \
  and\ \bibinfo {author} {\bibfnamefont {B.~K.}\ \bibnamefont {Agrawal}},\
  }\href {\doibase 10.1103/PhysRevC.76.045801} {\bibfield  {journal} {\bibinfo
  {journal} {Phys. Rev. C}\ }\textbf {\bibinfo {volume} {76}},\ \bibinfo
  {pages} {045801} (\bibinfo {year} {2007})}\BibitemShut {NoStop}%
\bibitem [{\citenamefont {Agrawal}(2010)}]{PhysRevC.81.034323}%
  \BibitemOpen
  \bibfield  {author} {\bibinfo {author} {\bibfnamefont {B.~K.}\ \bibnamefont
  {Agrawal}},\ }\href {\doibase 10.1103/PhysRevC.81.034323} {\bibfield
  {journal} {\bibinfo  {journal} {Phys. Rev. C}\ }\textbf {\bibinfo {volume}
  {81}},\ \bibinfo {pages} {034323} (\bibinfo {year} {2010})}\BibitemShut
  {NoStop}%
\bibitem [{\citenamefont {Fortin}\ \emph {et~al.}(2016)\citenamefont {Fortin},
  \citenamefont {Provid\^encia}, \citenamefont {Raduta}, \citenamefont
  {Gulminelli}, \citenamefont {Zdunik}, \citenamefont {Haensel},\ and\
  \citenamefont {Bejger}}]{PhysRevC.94.035804}%
  \BibitemOpen
  \bibfield  {author} {\bibinfo {author} {\bibfnamefont {M.}~\bibnamefont
  {Fortin}}, \bibinfo {author} {\bibfnamefont {C.}~\bibnamefont
  {Provid\^encia}}, \bibinfo {author} {\bibfnamefont {A.~R.}\ \bibnamefont
  {Raduta}}, \bibinfo {author} {\bibfnamefont {F.}~\bibnamefont {Gulminelli}},
  \bibinfo {author} {\bibfnamefont {J.~L.}\ \bibnamefont {Zdunik}}, \bibinfo
  {author} {\bibfnamefont {P.}~\bibnamefont {Haensel}}, \ and\ \bibinfo
  {author} {\bibfnamefont {M.}~\bibnamefont {Bejger}},\ }\href {\doibase
  10.1103/PhysRevC.94.035804} {\bibfield  {journal} {\bibinfo  {journal} {Phys.
  Rev. C}\ }\textbf {\bibinfo {volume} {94}},\ \bibinfo {pages} {035804}
  (\bibinfo {year} {2016})}\BibitemShut {NoStop}%
\bibitem [{\citenamefont {Whittenbury}\ \emph {et~al.}(2016)\citenamefont
  {Whittenbury}, \citenamefont {Matevosyan},\ and\ \citenamefont
  {Thomas}}]{PhysRevC.93.035807}%
  \BibitemOpen
  \bibfield  {author} {\bibinfo {author} {\bibfnamefont {D.~L.}\ \bibnamefont
  {Whittenbury}}, \bibinfo {author} {\bibfnamefont {H.~H.}\ \bibnamefont
  {Matevosyan}}, \ and\ \bibinfo {author} {\bibfnamefont {A.~W.}\ \bibnamefont
  {Thomas}},\ }\href {\doibase 10.1103/PhysRevC.93.035807} {\bibfield
  {journal} {\bibinfo  {journal} {Phys. Rev. C}\ }\textbf {\bibinfo {volume}
  {93}},\ \bibinfo {pages} {035807} (\bibinfo {year} {2016})}\BibitemShut
  {NoStop}%
\bibitem [{\citenamefont {Kojo}\ \emph {et~al.}(2016)\citenamefont {Kojo},
  \citenamefont {Powell}, \citenamefont {Song},\ and\ \citenamefont
  {Baym}}]{KOJO2016821}%
  \BibitemOpen
  \bibfield  {author} {\bibinfo {author} {\bibfnamefont {T.}~\bibnamefont
  {Kojo}}, \bibinfo {author} {\bibfnamefont {P.~D.}\ \bibnamefont {Powell}},
  \bibinfo {author} {\bibfnamefont {Y.}~\bibnamefont {Song}}, \ and\ \bibinfo
  {author} {\bibfnamefont {G.}~\bibnamefont {Baym}},\ }\href {\doibase
  https://doi.org/10.1016/j.nuclphysa.2016.02.008} {\bibfield  {journal}
  {\bibinfo  {journal} {Nucl. Phys. A}\ }\textbf {\bibinfo {volume} {956}},\
  \bibinfo {pages} {821 } (\bibinfo {year} {2016})},\ \bibinfo {note} {the XXV
  International Conference on Ultrarelativistic Nucleus-Nucleus Collisions:
  Quark Matter 2015}\BibitemShut {NoStop}%
\bibitem [{\citenamefont {Baym}\ \emph {et~al.}(1971)\citenamefont {Baym},
  \citenamefont {Pethick},\ and\ \citenamefont
  {Sutherland}}]{1971ApJ...170..299B}%
  \BibitemOpen
  \bibfield  {author} {\bibinfo {author} {\bibfnamefont {G.}~\bibnamefont
  {Baym}}, \bibinfo {author} {\bibfnamefont {C.}~\bibnamefont {Pethick}}, \
  and\ \bibinfo {author} {\bibfnamefont {P.}~\bibnamefont {Sutherland}},\
  }\href {\doibase 10.1086/151216} {\bibfield  {journal} {\bibinfo  {journal}
  {\apj}\ }\textbf {\bibinfo {volume} {170}},\ \bibinfo {pages} {299} (\bibinfo
  {year} {1971})}\BibitemShut {NoStop}%
\bibitem [{\citenamefont {Hansen}\ \emph {et~al.}(2007)\citenamefont {Hansen},
  \citenamefont {Alberico}, \citenamefont {Beraudo}, \citenamefont {Molinari},
  \citenamefont {Nardi},\ and\ \citenamefont {Ratti}}]{PhysRevD.75.065004}%
  \BibitemOpen
  \bibfield  {author} {\bibinfo {author} {\bibfnamefont {H.}~\bibnamefont
  {Hansen}}, \bibinfo {author} {\bibfnamefont {W.~M.}\ \bibnamefont
  {Alberico}}, \bibinfo {author} {\bibfnamefont {A.}~\bibnamefont {Beraudo}},
  \bibinfo {author} {\bibfnamefont {A.}~\bibnamefont {Molinari}}, \bibinfo
  {author} {\bibfnamefont {M.}~\bibnamefont {Nardi}}, \ and\ \bibinfo {author}
  {\bibfnamefont {C.}~\bibnamefont {Ratti}},\ }\href {\doibase
  10.1103/PhysRevD.75.065004} {\bibfield  {journal} {\bibinfo  {journal} {Phys.
  Rev. D}\ }\textbf {\bibinfo {volume} {75}},\ \bibinfo {pages} {065004}
  (\bibinfo {year} {2007})}\BibitemShut {NoStop}%
\bibitem [{\citenamefont {Costa}\ \emph {et~al.}(2010)\citenamefont {Costa},
  \citenamefont {Hansen}, \citenamefont {Ruivo},\ and\ \citenamefont
  {de~Sousa}}]{PhysRevD.81.016007}%
  \BibitemOpen
  \bibfield  {author} {\bibinfo {author} {\bibfnamefont {P.}~\bibnamefont
  {Costa}}, \bibinfo {author} {\bibfnamefont {H.}~\bibnamefont {Hansen}},
  \bibinfo {author} {\bibfnamefont {M.~C.}\ \bibnamefont {Ruivo}}, \ and\
  \bibinfo {author} {\bibfnamefont {C.~A.}\ \bibnamefont {de~Sousa}},\ }\href
  {\doibase 10.1103/PhysRevD.81.016007} {\bibfield  {journal} {\bibinfo
  {journal} {Phys. Rev. D}\ }\textbf {\bibinfo {volume} {81}},\ \bibinfo
  {pages} {016007} (\bibinfo {year} {2010})}\BibitemShut {NoStop}%
\bibitem [{\citenamefont {MOREIRA}\ \emph {et~al.}(2012)\citenamefont
  {MOREIRA}, \citenamefont {HILLER}, \citenamefont {OSIPOV},\ and\
  \citenamefont {BLIN}}]{doi:10.1142/S0217751X12500601}%
  \BibitemOpen
  \bibfield  {author} {\bibinfo {author} {\bibfnamefont {J.}~\bibnamefont
  {MOREIRA}}, \bibinfo {author} {\bibfnamefont {B.}~\bibnamefont {HILLER}},
  \bibinfo {author} {\bibfnamefont {A.~A.}\ \bibnamefont {OSIPOV}}, \ and\
  \bibinfo {author} {\bibfnamefont {A.~H.}\ \bibnamefont {BLIN}},\ }\href
  {\doibase 10.1142/S0217751X12500601} {\bibfield  {journal} {\bibinfo
  {journal} {Int. J. Mod. Phys. A}\ }\textbf {\bibinfo {volume} {27}},\
  \bibinfo {pages} {1250060} (\bibinfo {year} {2012})}\BibitemShut {NoStop}%
\bibitem [{\citenamefont {Bratovic}\ \emph {et~al.}(2013)\citenamefont
  {Bratovic}, \citenamefont {Hatsuda},\ and\ \citenamefont
  {Weise}}]{BRATOVIC2013131}%
  \BibitemOpen
  \bibfield  {author} {\bibinfo {author} {\bibfnamefont {N.}~\bibnamefont
  {Bratovic}}, \bibinfo {author} {\bibfnamefont {T.}~\bibnamefont {Hatsuda}}, \
  and\ \bibinfo {author} {\bibfnamefont {W.}~\bibnamefont {Weise}},\ }\href
  {\doibase http://dx.doi.org/10.1016/j.physletb.2013.01.003} {\bibfield
  {journal} {\bibinfo  {journal} {Phys. Lett. B}\ }\textbf {\bibinfo {volume}
  {719}},\ \bibinfo {pages} {131 } (\bibinfo {year} {2013})}\BibitemShut
  {NoStop}%
\bibitem [{\citenamefont {Fukushima}(2004)}]{FUKUSHIMA2004277}%
  \BibitemOpen
  \bibfield  {author} {\bibinfo {author} {\bibfnamefont {K.}~\bibnamefont
  {Fukushima}},\ }\href {\doibase
  http://dx.doi.org/10.1016/j.physletb.2004.04.027} {\bibfield  {journal}
  {\bibinfo  {journal} {Phys. Lett. B}\ }\textbf {\bibinfo {volume} {591}},\
  \bibinfo {pages} {277 } (\bibinfo {year} {2004})}\BibitemShut {NoStop}%
\bibitem [{\citenamefont {Ruggieri}(2011)}]{PhysRevD.84.014011}%
  \BibitemOpen
  \bibfield  {author} {\bibinfo {author} {\bibfnamefont {M.}~\bibnamefont
  {Ruggieri}},\ }\href {\doibase 10.1103/PhysRevD.84.014011} {\bibfield
  {journal} {\bibinfo  {journal} {Phys. Rev. D}\ }\textbf {\bibinfo {volume}
  {84}},\ \bibinfo {pages} {014011} (\bibinfo {year} {2011})}\BibitemShut
  {NoStop}%
\bibitem [{\citenamefont {Chernodub}\ and\ \citenamefont
  {Nedelin}(2011)}]{PhysRevD.83.105008}%
  \BibitemOpen
  \bibfield  {author} {\bibinfo {author} {\bibfnamefont {M.~N.}\ \bibnamefont
  {Chernodub}}\ and\ \bibinfo {author} {\bibfnamefont {A.~S.}\ \bibnamefont
  {Nedelin}},\ }\href {\doibase 10.1103/PhysRevD.83.105008} {\bibfield
  {journal} {\bibinfo  {journal} {Phys. Rev. D}\ }\textbf {\bibinfo {volume}
  {83}},\ \bibinfo {pages} {105008} (\bibinfo {year} {2011})}\BibitemShut
  {NoStop}%
\bibitem [{\citenamefont {Gatto}\ and\ \citenamefont
  {Ruggieri}(2012)}]{PhysRevD.85.054013}%
  \BibitemOpen
  \bibfield  {author} {\bibinfo {author} {\bibfnamefont {R.}~\bibnamefont
  {Gatto}}\ and\ \bibinfo {author} {\bibfnamefont {M.}~\bibnamefont
  {Ruggieri}},\ }\href {\doibase 10.1103/PhysRevD.85.054013} {\bibfield
  {journal} {\bibinfo  {journal} {Phys. Rev. D}\ }\textbf {\bibinfo {volume}
  {85}},\ \bibinfo {pages} {054013} (\bibinfo {year} {2012})}\BibitemShut
  {NoStop}%
\bibitem [{\citenamefont {Roessner}\ \emph {et~al.}(2007)\citenamefont
  {Roessner}, \citenamefont {Ratti},\ and\ \citenamefont
  {Weise}}]{PhysRevD.75.034007}%
  \BibitemOpen
  \bibfield  {author} {\bibinfo {author} {\bibfnamefont {S.}~\bibnamefont
  {Roessner}}, \bibinfo {author} {\bibfnamefont {C.}~\bibnamefont {Ratti}}, \
  and\ \bibinfo {author} {\bibfnamefont {W.}~\bibnamefont {Weise}},\ }\href
  {\doibase 10.1103/PhysRevD.75.034007} {\bibfield  {journal} {\bibinfo
  {journal} {Phys. Rev. D}\ }\textbf {\bibinfo {volume} {75}},\ \bibinfo
  {pages} {034007} (\bibinfo {year} {2007})}\BibitemShut {NoStop}%
\bibitem [{\citenamefont {Cui}\ \emph {et~al.}(2016)\citenamefont {Cui},
  \citenamefont {Clo\"et}, \citenamefont {Lu}, \citenamefont {Roberts},
  \citenamefont {Schmidt}, \citenamefont {Xu},\ and\ \citenamefont
  {Zong}}]{PhysRevD.94.071503}%
  \BibitemOpen
  \bibfield  {author} {\bibinfo {author} {\bibfnamefont {Z.-F.}\ \bibnamefont
  {Cui}}, \bibinfo {author} {\bibfnamefont {I.~C.}\ \bibnamefont {Clo\"et}},
  \bibinfo {author} {\bibfnamefont {Y.}~\bibnamefont {Lu}}, \bibinfo {author}
  {\bibfnamefont {C.~D.}\ \bibnamefont {Roberts}}, \bibinfo {author}
  {\bibfnamefont {S.~M.}\ \bibnamefont {Schmidt}}, \bibinfo {author}
  {\bibfnamefont {S.-S.}\ \bibnamefont {Xu}}, \ and\ \bibinfo {author}
  {\bibfnamefont {H.-S.}\ \bibnamefont {Zong}},\ }\href {\doibase
  10.1103/PhysRevD.94.071503} {\bibfield  {journal} {\bibinfo  {journal} {Phys.
  Rev. D}\ }\textbf {\bibinfo {volume} {94}},\ \bibinfo {pages} {071503}
  (\bibinfo {year} {2016})}\BibitemShut {NoStop}%
\bibitem [{\citenamefont {Hatsuda}\ and\ \citenamefont
  {Kunihiro}(1994)}]{HATSUDA1994221}%
  \BibitemOpen
  \bibfield  {author} {\bibinfo {author} {\bibfnamefont {T.}~\bibnamefont
  {Hatsuda}}\ and\ \bibinfo {author} {\bibfnamefont {T.}~\bibnamefont
  {Kunihiro}},\ }\href {\doibase
  http://dx.doi.org/10.1016/0370-1573(94)90022-1} {\bibfield  {journal}
  {\bibinfo  {journal} {Phys. Rep.}\ }\textbf {\bibinfo {volume} {247}},\
  \bibinfo {pages} {221 } (\bibinfo {year} {1994})}\BibitemShut {NoStop}%
\bibitem [{\citenamefont {Fischer}\ and\ \citenamefont
  {Luecker}(2013)}]{FISCHER20131036}%
  \BibitemOpen
  \bibfield  {author} {\bibinfo {author} {\bibfnamefont {C.~S.}\ \bibnamefont
  {Fischer}}\ and\ \bibinfo {author} {\bibfnamefont {J.}~\bibnamefont
  {Luecker}},\ }\href {\doibase https://doi.org/10.1016/j.physletb.2012.11.054}
  {\bibfield  {journal} {\bibinfo  {journal} {Phys. Lett. B}\ }\textbf
  {\bibinfo {volume} {718}},\ \bibinfo {pages} {1036 } (\bibinfo {year}
  {2013})}\BibitemShut {NoStop}%
\bibitem [{\citenamefont {Fischer}\ \emph {et~al.}(2011)\citenamefont
  {Fischer}, \citenamefont {Luecker},\ and\ \citenamefont
  {Mueller}}]{FISCHER2011438}%
  \BibitemOpen
  \bibfield  {author} {\bibinfo {author} {\bibfnamefont {C.~S.}\ \bibnamefont
  {Fischer}}, \bibinfo {author} {\bibfnamefont {J.}~\bibnamefont {Luecker}}, \
  and\ \bibinfo {author} {\bibfnamefont {J.~A.}\ \bibnamefont {Mueller}},\
  }\href {\doibase https://doi.org/10.1016/j.physletb.2011.07.039} {\bibfield
  {journal} {\bibinfo  {journal} {Phys. Lett. B}\ }\textbf {\bibinfo {volume}
  {702}},\ \bibinfo {pages} {438 } (\bibinfo {year} {2011})}\BibitemShut
  {NoStop}%
\bibitem [{\citenamefont {Bors{\'a}nyi}\ \emph {et~al.}(2010)\citenamefont
  {Bors{\'a}nyi}, \citenamefont {Fodor}, \citenamefont {Hoelbling},
  \citenamefont {Katz}, \citenamefont {Krieg}, \citenamefont {Ratti},\ and\
  \citenamefont {Szab{\'o}}}]{Borsanyi2010}%
  \BibitemOpen
  \bibfield  {author} {\bibinfo {author} {\bibfnamefont {S.}~\bibnamefont
  {Bors{\'a}nyi}}, \bibinfo {author} {\bibfnamefont {Z.}~\bibnamefont {Fodor}},
  \bibinfo {author} {\bibfnamefont {C.}~\bibnamefont {Hoelbling}}, \bibinfo
  {author} {\bibfnamefont {S.~D.}\ \bibnamefont {Katz}}, \bibinfo {author}
  {\bibfnamefont {S.}~\bibnamefont {Krieg}}, \bibinfo {author} {\bibfnamefont
  {C.}~\bibnamefont {Ratti}}, \ and\ \bibinfo {author} {\bibfnamefont {K.~K.}\
  \bibnamefont {Szab{\'o}}},\ }\href {\doibase 10.1007/JHEP09(2010)073}
  {\bibfield  {journal} {\bibinfo  {journal} {J. High Energy Phys.}\ }\textbf
  {\bibinfo {volume} {2010}},\ \bibinfo {pages} {73} (\bibinfo {year}
  {2010})}\BibitemShut {NoStop}%
\bibitem [{\citenamefont {Hiller}\ \emph {et~al.}(2010)\citenamefont {Hiller},
  \citenamefont {Moreira}, \citenamefont {Osipov},\ and\ \citenamefont
  {Blin}}]{PhysRevD.81.116005}%
  \BibitemOpen
  \bibfield  {author} {\bibinfo {author} {\bibfnamefont {B.}~\bibnamefont
  {Hiller}}, \bibinfo {author} {\bibfnamefont {J.}~\bibnamefont {Moreira}},
  \bibinfo {author} {\bibfnamefont {A.~A.}\ \bibnamefont {Osipov}}, \ and\
  \bibinfo {author} {\bibfnamefont {A.~H.}\ \bibnamefont {Blin}},\ }\href
  {\doibase 10.1103/PhysRevD.81.116005} {\bibfield  {journal} {\bibinfo
  {journal} {Phys. Rev. D}\ }\textbf {\bibinfo {volume} {81}},\ \bibinfo
  {pages} {116005} (\bibinfo {year} {2010})}\BibitemShut {NoStop}%
\bibitem [{\citenamefont {Osipov}\ \emph {et~al.}(2006)\citenamefont {Osipov},
  \citenamefont {Hiller},\ and\ \citenamefont {da~Providencia}}]{OSIPOV200648}%
  \BibitemOpen
  \bibfield  {author} {\bibinfo {author} {\bibfnamefont {A.}~\bibnamefont
  {Osipov}}, \bibinfo {author} {\bibfnamefont {B.}~\bibnamefont {Hiller}}, \
  and\ \bibinfo {author} {\bibfnamefont {J.}~\bibnamefont {da~Providencia}},\
  }\href {\doibase https://doi.org/10.1016/j.physletb.2006.01.008} {\bibfield
  {journal} {\bibinfo  {journal} {Phys. Lett. B}\ }\textbf {\bibinfo {volume}
  {634}},\ \bibinfo {pages} {48 } (\bibinfo {year} {2006})}\BibitemShut
  {NoStop}%
\bibitem [{\citenamefont {Osipov}\ \emph {et~al.}(2007)\citenamefont {Osipov},
  \citenamefont {Hiller}, \citenamefont {Blin},\ and\ \citenamefont
  {da~Providencia}}]{OSIPOV20072021}%
  \BibitemOpen
  \bibfield  {author} {\bibinfo {author} {\bibfnamefont {A.}~\bibnamefont
  {Osipov}}, \bibinfo {author} {\bibfnamefont {B.}~\bibnamefont {Hiller}},
  \bibinfo {author} {\bibfnamefont {A.}~\bibnamefont {Blin}}, \ and\ \bibinfo
  {author} {\bibfnamefont {J.}~\bibnamefont {da~Providencia}},\ }\href
  {\doibase https://doi.org/10.1016/j.aop.2006.08.004} {\bibfield  {journal}
  {\bibinfo  {journal} {Ann. Phys.}\ }\textbf {\bibinfo {volume} {322}},\
  \bibinfo {pages} {2021 } (\bibinfo {year} {2007})}\BibitemShut {NoStop}%
\bibitem [{\citenamefont {Fukushima}\ and\ \citenamefont
  {Hatsuda}(2011)}]{0034-4885-74-1-014001}%
  \BibitemOpen
  \bibfield  {author} {\bibinfo {author} {\bibfnamefont {K.}~\bibnamefont
  {Fukushima}}\ and\ \bibinfo {author} {\bibfnamefont {T.}~\bibnamefont
  {Hatsuda}},\ }\href {http://stacks.iop.org/0034-4885/74/i=1/a=014001}
  {\bibfield  {journal} {\bibinfo  {journal} {Rep. Prog. Phys.}\ }\textbf
  {\bibinfo {volume} {74}},\ \bibinfo {pages} {014001} (\bibinfo {year}
  {2011})}\BibitemShut {NoStop}%
\bibitem [{\citenamefont {ZONG}\ and\ \citenamefont
  {SUN}(2008)}]{doi:10.1142/S0217751X08040457}%
  \BibitemOpen
  \bibfield  {author} {\bibinfo {author} {\bibfnamefont {H.-S.}\ \bibnamefont
  {ZONG}}\ and\ \bibinfo {author} {\bibfnamefont {W.-M.}\ \bibnamefont {SUN}},\
  }\href {\doibase 10.1142/S0217751X08040457} {\bibfield  {journal} {\bibinfo
  {journal} {Int. J. Mod. Phys. A}\ }\textbf {\bibinfo {volume} {23}},\
  \bibinfo {pages} {3591} (\bibinfo {year} {2008})}\BibitemShut {NoStop}%
\bibitem [{\citenamefont {Song}\ \emph {et~al.}(1992)\citenamefont {Song},
  \citenamefont {Enke},\ and\ \citenamefont {Jiarong}}]{PhysRevD.46.3211}%
  \BibitemOpen
  \bibfield  {author} {\bibinfo {author} {\bibfnamefont {G.}~\bibnamefont
  {Song}}, \bibinfo {author} {\bibfnamefont {W.}~\bibnamefont {Enke}}, \ and\
  \bibinfo {author} {\bibfnamefont {L.}~\bibnamefont {Jiarong}},\ }\href
  {\doibase 10.1103/PhysRevD.46.3211} {\bibfield  {journal} {\bibinfo
  {journal} {Phys. Rev. D}\ }\textbf {\bibinfo {volume} {46}},\ \bibinfo
  {pages} {3211} (\bibinfo {year} {1992})}\BibitemShut {NoStop}%
\bibitem [{\citenamefont {Lu}\ \emph {et~al.}(1998)\citenamefont {Lu},
  \citenamefont {Tsushima}, \citenamefont {Thomas}, \citenamefont {Williams},\
  and\ \citenamefont {Saito}}]{LU1998443}%
  \BibitemOpen
  \bibfield  {author} {\bibinfo {author} {\bibfnamefont {D.}~\bibnamefont
  {Lu}}, \bibinfo {author} {\bibfnamefont {K.}~\bibnamefont {Tsushima}},
  \bibinfo {author} {\bibfnamefont {A.}~\bibnamefont {Thomas}}, \bibinfo
  {author} {\bibfnamefont {A.}~\bibnamefont {Williams}}, \ and\ \bibinfo
  {author} {\bibfnamefont {K.}~\bibnamefont {Saito}},\ }\href {\doibase
  https://doi.org/10.1016/S0375-9474(98)00181-X} {\bibfield  {journal}
  {\bibinfo  {journal} {Nucl. Phys. A}\ }\textbf {\bibinfo {volume} {634}},\
  \bibinfo {pages} {443 } (\bibinfo {year} {1998})}\BibitemShut {NoStop}%
\bibitem [{\citenamefont {Zhou}\ \emph {et~al.}(2017)\citenamefont {Zhou},
  \citenamefont {Zhou},\ and\ \citenamefont {Li}}]{Zhou:2017pha}%
  \BibitemOpen
  \bibfield  {author} {\bibinfo {author} {\bibfnamefont {E.-P.}\ \bibnamefont
  {Zhou}}, \bibinfo {author} {\bibfnamefont {X.}~\bibnamefont {Zhou}}, \ and\
  \bibinfo {author} {\bibfnamefont {A.}~\bibnamefont {Li}},\ }\href@noop {} {\
  (\bibinfo {year} {2017})},\ \Eprint {http://arxiv.org/abs/1711.04312}
  {arXiv:1711.04312 [astro-ph.HE]} \BibitemShut {NoStop}%
\bibitem [{\citenamefont {Yan}\ \emph {et~al.}(2012)\citenamefont {Yan},
  \citenamefont {Cao}, \citenamefont {Luo}, \citenamefont {Sun},\ and\
  \citenamefont {Zong}}]{PhysRevD.86.114028}%
  \BibitemOpen
  \bibfield  {author} {\bibinfo {author} {\bibfnamefont {Y.}~\bibnamefont
  {Yan}}, \bibinfo {author} {\bibfnamefont {J.}~\bibnamefont {Cao}}, \bibinfo
  {author} {\bibfnamefont {X.-L.}\ \bibnamefont {Luo}}, \bibinfo {author}
  {\bibfnamefont {W.-M.}\ \bibnamefont {Sun}}, \ and\ \bibinfo {author}
  {\bibfnamefont {H.}~\bibnamefont {Zong}},\ }\href {\doibase
  10.1103/PhysRevD.86.114028} {\bibfield  {journal} {\bibinfo  {journal} {Phys.
  Rev. D}\ }\textbf {\bibinfo {volume} {86}},\ \bibinfo {pages} {114028}
  (\bibinfo {year} {2012})}\BibitemShut {NoStop}%
\bibitem [{\citenamefont {Benvenuto}\ and\ \citenamefont
  {Lugones}(1995)}]{PhysRevD.51.1989}%
  \BibitemOpen
  \bibfield  {author} {\bibinfo {author} {\bibfnamefont {O.~G.}\ \bibnamefont
  {Benvenuto}}\ and\ \bibinfo {author} {\bibfnamefont {G.}~\bibnamefont
  {Lugones}},\ }\href {\doibase 10.1103/PhysRevD.51.1989} {\bibfield  {journal}
  {\bibinfo  {journal} {Phys. Rev. D}\ }\textbf {\bibinfo {volume} {51}},\
  \bibinfo {pages} {1989} (\bibinfo {year} {1995})}\BibitemShut {NoStop}%
\bibitem [{\citenamefont {Schertler}\ \emph {et~al.}(1999)\citenamefont
  {Schertler}, \citenamefont {Leupold},\ and\ \citenamefont
  {Schaffner-Bielich}}]{PhysRevC.60.025801}%
  \BibitemOpen
  \bibfield  {author} {\bibinfo {author} {\bibfnamefont {K.}~\bibnamefont
  {Schertler}}, \bibinfo {author} {\bibfnamefont {S.}~\bibnamefont {Leupold}},
  \ and\ \bibinfo {author} {\bibfnamefont {J.}~\bibnamefont
  {Schaffner-Bielich}},\ }\href {\doibase 10.1103/PhysRevC.60.025801}
  {\bibfield  {journal} {\bibinfo  {journal} {Phys. Rev. C}\ }\textbf {\bibinfo
  {volume} {60}},\ \bibinfo {pages} {025801} (\bibinfo {year}
  {1999})}\BibitemShut {NoStop}%
\bibitem [{\citenamefont {Hoyos}\ \emph {et~al.}(2016)\citenamefont {Hoyos},
  \citenamefont {Jokela}, \citenamefont {Rodr\'{\i}guez~Fern\'andez},\ and\
  \citenamefont {Vuorinen}}]{PhysRevLett.117.032501}%
  \BibitemOpen
  \bibfield  {author} {\bibinfo {author} {\bibfnamefont {C.}~\bibnamefont
  {Hoyos}}, \bibinfo {author} {\bibfnamefont {N.}~\bibnamefont {Jokela}},
  \bibinfo {author} {\bibfnamefont {D.}~\bibnamefont
  {Rodr\'{\i}guez~Fern\'andez}}, \ and\ \bibinfo {author} {\bibfnamefont
  {A.}~\bibnamefont {Vuorinen}},\ }\href {\doibase
  10.1103/PhysRevLett.117.032501} {\bibfield  {journal} {\bibinfo  {journal}
  {Phys. Rev. Lett.}\ }\textbf {\bibinfo {volume} {117}},\ \bibinfo {pages}
  {032501} (\bibinfo {year} {2016})}\BibitemShut {NoStop}%
\bibitem [{\citenamefont {Shifman}(1998)}]{doi:10.1143/PTPS.131.1}%
  \BibitemOpen
  \bibfield  {author} {\bibinfo {author} {\bibfnamefont {M.}~\bibnamefont
  {Shifman}},\ }\href {\doibase 10.1143/PTPS.131.1} {\bibfield  {journal}
  {\bibinfo  {journal} {Prog. Theor. Phys. Suppl.}\ }\textbf {\bibinfo {volume}
  {131}},\ \bibinfo {pages} {1} (\bibinfo {year} {1998})}\BibitemShut {NoStop}%
\bibitem [{\citenamefont {McNeile}\ \emph {et~al.}(2013)\citenamefont
  {McNeile}, \citenamefont {Bazavov}, \citenamefont {Davies}, \citenamefont
  {Dowdall}, \citenamefont {Hornbostel}, \citenamefont {Lepage},\ and\
  \citenamefont {Trottier}}]{PhysRevD.87.034503}%
  \BibitemOpen
  \bibfield  {author} {\bibinfo {author} {\bibfnamefont {C.}~\bibnamefont
  {McNeile}}, \bibinfo {author} {\bibfnamefont {A.}~\bibnamefont {Bazavov}},
  \bibinfo {author} {\bibfnamefont {C.~T.~H.}\ \bibnamefont {Davies}}, \bibinfo
  {author} {\bibfnamefont {R.~J.}\ \bibnamefont {Dowdall}}, \bibinfo {author}
  {\bibfnamefont {K.}~\bibnamefont {Hornbostel}}, \bibinfo {author}
  {\bibfnamefont {G.~P.}\ \bibnamefont {Lepage}}, \ and\ \bibinfo {author}
  {\bibfnamefont {H.~D.}\ \bibnamefont {Trottier}},\ }\href {\doibase
  10.1103/PhysRevD.87.034503} {\bibfield  {journal} {\bibinfo  {journal} {Phys.
  Rev. D}\ }\textbf {\bibinfo {volume} {87}},\ \bibinfo {pages} {034503}
  (\bibinfo {year} {2013})}\BibitemShut {NoStop}%
\bibitem [{\citenamefont {Maris}\ and\ \citenamefont
  {Roberts}(1997)}]{PhysRevC.56.3369}%
  \BibitemOpen
  \bibfield  {author} {\bibinfo {author} {\bibfnamefont {P.}~\bibnamefont
  {Maris}}\ and\ \bibinfo {author} {\bibfnamefont {C.~D.}\ \bibnamefont
  {Roberts}},\ }\href {\doibase 10.1103/PhysRevC.56.3369} {\bibfield  {journal}
  {\bibinfo  {journal} {Phys. Rev. C}\ }\textbf {\bibinfo {volume} {56}},\
  \bibinfo {pages} {3369} (\bibinfo {year} {1997})}\BibitemShut {NoStop}%
\bibitem [{\citenamefont {Zong}\ \emph {et~al.}(2003)\citenamefont {Zong},
  \citenamefont {Ping}, \citenamefont {Yang}, \citenamefont {L\"u},\ and\
  \citenamefont {Wang}}]{PhysRevD.67.074004}%
  \BibitemOpen
  \bibfield  {author} {\bibinfo {author} {\bibfnamefont {H.-s.}\ \bibnamefont
  {Zong}}, \bibinfo {author} {\bibfnamefont {J.-l.}\ \bibnamefont {Ping}},
  \bibinfo {author} {\bibfnamefont {H.-t.}\ \bibnamefont {Yang}}, \bibinfo
  {author} {\bibfnamefont {X.-f.}\ \bibnamefont {L\"u}}, \ and\ \bibinfo
  {author} {\bibfnamefont {F.}~\bibnamefont {Wang}},\ }\href {\doibase
  10.1103/PhysRevD.67.074004} {\bibfield  {journal} {\bibinfo  {journal} {Phys.
  Rev. D}\ }\textbf {\bibinfo {volume} {67}},\ \bibinfo {pages} {074004}
  (\bibinfo {year} {2003})}\BibitemShut {NoStop}%
\end{thebibliography}%
\end{document}